\documentclass[twocolumn,preprintnumbers,nofootinbib,prd,superscriptaddress,aps]{revtex4-1}

\usepackage[utf8]{inputenc}
\usepackage{graphicx,amssymb,amsmath,amsthm,amsfonts,epstopdf,epsfig,epsf,times}
\usepackage[linktocpage]{hyperref}
\usepackage[usenames]{color}
\usepackage{epstopdf}

\usepackage{aas_macros}
\usepackage{bm}
\usepackage{dcolumn}
\usepackage{latexsym}
\usepackage{rotating}
\usepackage{hyperref}
\usepackage{color}
\usepackage{longtable}
\usepackage{enumerate}
\usepackage{tensor}
\usepackage{stmaryrd}

\usepackage{mathtools}
\usepackage{url}
\definecolor{coolblack}{rgb}{0.0, 0.18, 0.39}
\definecolor{darkred}{rgb}{0.5,0,0}
\definecolor{darkgreen}{rgb}{0,0.5,0}
\definecolor{darkblue}{rgb}{0,0,0.5}
\definecolor{lapislazuli}{rgb}{0.15, 0.38, 0.61}
\definecolor{venetianred}{rgb}{0.78, 0.03, 0.08}
\definecolor{bleudefrance}{rgb}{0.19, 0.55, 0.91}
\definecolor{dogwoodrose}{rgb}{0.84, 0.09, 0.41}
\definecolor{dogwoodrose}{rgb}{0.84, 0.09, 0.41}
\hypersetup{colorlinks=true, citecolor=darkblue, linkcolor=darkblue, 
urlcolor = darkblue}

\definecolor{olive}{rgb}{0.5, 0.5, 0.0}

\def\nn{\nonumber}

\newcommand{\ben}{\begin{enumerate}}
\newcommand{\een}{\end{enumerate}}

\def\be{\begin{equation}}
\def\ee{\end{equation}}
\newcommand{\beq}{\begin{eqnarray}}
\newcommand{\eeq}{\end{eqnarray}} 
\newcommand{\ba}{\begin{align}}
\newcommand{\ea}{\end{align}}

\def\be{\begin{equation}}
\def\ee{\end{equation}}
\def\p{\partial}	
\newcommand{\bea}{\begin{eqnarray}}
\newcommand{\eea}{\end{eqnarray}}
\def\l{\left}
\def\r{\right}

\begin{document}\title {\large Orbital fingerprints of ultralight scalar fields around black holes}

\author{Miguel C. Ferreira}
\affiliation{CENTRA, Departamento de F\'{\i}sica, Instituto Superior T\'ecnico -- IST, Universidade de Lisboa -- UL,
Avenida Rovisco Pais 1, 1049 Lisboa, Portugal}

\author{Caio F. B. Macedo}
\affiliation{CENTRA, Departamento de F\'{\i}sica, Instituto Superior T\'ecnico -- IST, Universidade de Lisboa -- UL,
Avenida Rovisco Pais 1, 1049 Lisboa, Portugal}
\affiliation{Faculdade de F\'{\i}sica, Universidade 
Federal do Par\'a, 66075-110, Bel\'em, Par\'a, Brazil.}
\affiliation{Campus Salinópolis, Universidade 
Federal do Par\'a, 68721-000, Salinópolis, Par\'a, Brazil.}
\author{Vitor Cardoso}
\affiliation{CENTRA, Departamento de F\'{\i}sica, Instituto Superior T\'ecnico -- IST, Universidade de Lisboa -- UL,
Avenida Rovisco Pais 1, 1049 Lisboa, Portugal}
\affiliation{Perimeter Institute for Theoretical Physics, 31 Caroline Street North
Waterloo, Ontario N2L 2Y5, Canada}

\begin{abstract}
Ultralight scalars have been predicted in a variety of scenarios, and advocated as a possible component of dark matter. These fields can form compact regular structures known as boson stars, or---in the presence of horizons---give rise to nontrivial time-dependent scalar hair and a stationary geometry. Because these fields can be coherent over large spatial extents, their interaction with ``regular'' matter can lead to very peculiar effects, most notably resonances. Here we study the motion of stars in a background describing black holes surrounded by non-axially symmetric scalar field profiles. By analyzing the system in a weak-field approach, we find that the presence of a scalar field gives rise to secular effects akin to ones existing in planetary and accretion disks. Particularly, the existence of resonances between the orbiting stars and the scalar field may enable angular momentum exchange between them, providing mechanisms similar to planetary migration. Additionally, these mechanisms may allow \textit{floating orbits}, which are stable radiating orbits. We also show, in the full relativistic case, that these effects also appear when there is a direct coupling between the scalar field and the stellar matter, which can arise due to the presence of a scalar core in the star or in alternative theories of gravity.
\end{abstract}

\date{\today}

\maketitle

\tableofcontents

\section{Introduction}\label{sec:intro}

\subsection{Motivation}

Scalar fields are ubiquitous models to describe complex phenomena, and are frequently used as effective descriptions or to capture what is thought to be the essentials of new interactions.
The examples are too many to describe in any useful detail, and include models of dark matter halos \cite{Hui:2016ltb,Urena-Lopez:2017tob}, fields assisting inflation during the early stages of the universe, or even strong-gravity effects such as scalarization in the interior of compact stars~\cite{Berti:2015itd}.

Among the consequences of the existence of scalar fields are self-gravitating structures, like boson stars and oscillatons which exist when the field is complex or real, respectively. These structures have been used extensively as models for dark matter and of compact objects~\cite{Kaup:1968zz,Ruffini:1969qy,Jetzer:1991jr,Schunck:2003kk,Liebling:2012fv,Macedo:2013jja,Brito:2015pxa,Seidel:1991zh,Okawa:2013jba}, usually requiring the scalar field to be massive (in order not to disperse to infinity), and time-dependent (in order to create enough pressure to sustain from collapsing). 
Massive scalar {\it real} fields around black holes (BHs) or collapsing stars can lead to very long-lived---for all purposes stationary---configurations~\cite{Witek:2012tr,Okawa:2014nda,Barranco:2012qs,Sanchis-Gual:2014ewa,Sanchis-Gual:2015sxa}, while {\it complex} fields may form truly stationary configurations~\cite{Herdeiro:2014goa,Herdeiro:2015waa,Hod:2012px,Benone:2014ssa}, also dubbed as \textit{clouds}, which are kept from being absorbed by the horizon through a process known as superradiance~\cite{Brito:2015oca}. In fact, these hairy BH solutions are smoothly connected to spinning boson stars, such that they can be thought of as a spinning boson star, at the center of which a spinning BH (with carefully designed angular velocity) was placed. Notwithstanding, all of the above features can be generalize to massive vector fields instead of scalar ones~\cite{Witek:2012tr,Brito:2015yga,Brito:2015pxa,Herdeiro:2016tmi}.

Structures such as boson stars or scalar-hairy BHs are usually compact, and therefore apt to emit copious amounts of gravitational waves upon collisions or other interactions, which could in turn be used as a tool to discriminate them. Among some of the smoking-gun effects for long-range scalars, it was found that

\noindent {\bf (i)} BHs should have ``holes'' in the spin-mass plane (also known as Regge-plane), corresponding to the regions where the superradiant instability is effective~\cite{Brito:2015oca,Brito:2014wla,Arvanitaki:2016qwi,Arvanitaki:2014wva,Brito:2017wnc,Brito:2017zvb}.
Thus, observations of BHs and accurate estimates of their mass and spin could provide clear indications of the existence of light fields.

\noindent {\bf (ii)} Single BHs can act as sources of monochromatic gravitational waves, potentially detectable by LIGO or LISA, either as resolved events or as a stochastic background~\cite{Brito:2015oca,Brito:2014wla,Arvanitaki:2016qwi,Arvanitaki:2014wva,Brito:2017wnc,Brito:2017zvb}

\noindent {\bf (iii)} Stars or planets carrying scalar charge, or otherwise interacting non-minimally with it,
{\it will} probe resonances in the spacetime, where energy extraction from the horizon compensates for losses through gravitational radiation: the orbiting object floats at a fixed angular velocity for large timescales~\cite{Cardoso:2011xi,Fujita:2016yav}. The imprint on gravitational waves may be detectable~\cite{Yunes:2011aa}.

In addition, possible signs of scalar fields might be imprinted in the way that they affect the bending of light, i.e., in their shadows~\cite{Vincent:2015xta,Cunha:2015yba,Vincent:2016sjq} or in $X-$ray reflection spectrum from surrounding accretion disks~\cite{Zhou:2017glv}. Finally, any kind of perturbation around compact objects is likely to excite proper oscillation modes (the quasinormal modes), which can also be used to test the nature of the object~\cite{Cardoso:2016ryw}. This program however requires detailed knowledge of the response of the scalar field structure to external perturbations, which is still lacking.

Intuitively, it is expected that a spinning BH onto which a scalar field is ``fastened'' will drag the scalar field, and neighboring matter as well. In other words, there should be correlations between the characteristics of the scalar field and the behavior of stars and planets in its immediate neighborhood. We shall look for these correlations in Extreme-Mass-Ratio-Inspirals (EMRIs)---binary systems composed of a central Super Massive Black Hole (SMBH) orbited by a much lighter object (a white dwarf, neutron star or solar mass BH)---which constitute one of the most promising sources of gravitational radiation to be analyzed by upcoming facilities \cite{amaro-seoane2007,gair_living_review,Audley:2017drz}. By considering that the SMBH supports a non-axially symmetric scalar field configuration, we explore the gravitational effects that the presence of the scalar field imposes on the motion of the orbiting body. We argue that the angular momentum imparted in the orbiting body by the scalar field may balance the angular momentum lost by gravitational radiation, given an additional mechanism to enable floating orbits~\cite{Press:1972zz,Cardoso:2011xi}. In addition, we also take into account more speculative channels of interaction between the orbiting body and the scalar field, namely the possibility of it having a scalar charge or of being acted by friction forces. The final aim of the study we initiate here is to fully understand the theoretical aspects of the dynamics of EMRIs in order to use them as probes to the existence of long-lived scalar field configurations surrounding SMBH.

\subsection{Summary}
In what follows we focus on an EMRI in which the SMBH supports an ultralight scalar field---we shall abbreviate it to Black Hole-Scalar Field system (BHSFS). We study the impact of the scalar field on the orbital structure of the orbiting body with special emphasis on circular orbits. Our main findings are:
\begin{itemize}
  \item General orbits of the EMRI precess at a rate that depends on the parameters governing the scalar field. This effect adds to the precession caused by general relativistic terms and, therefore, could be probed by measuring carefully any additional amount in rates of precession of satellites;
  \item The existence of the scalar field configuration around the BH gives rise to resonant orbits when its rotation frequency is equal to one of the characteristic frequencies of the system; there are three resonances:
  \begin{itemize}
    \item One corotation resonance where, in the absence of other perturbing effects, large numbers of orbiting bodies (such as stars) will tend to pile up at the resonant radius;
    \item Two Lindblad (inner and outer) resonances where, at first order, the scalar field can exchange angular momentum with the orbiting particle,  an effect which is associated with orbital migration.
  \end{itemize}
\end{itemize}

The BHSFS system is introduced with detail in Sec.~\ref{sec:framework}, where we discuss the weak field limit and the parameters used in the description of the system. We analyze the stellar orbits around the BHSFS in Sec.~\ref{sec:quasicircular}. We also discuss the possibility of adding a non-minimal coupling and friction forces to the system in Sec.~\ref{sec:forces}. Finally, Sec.~\ref{sec:discusscf} fixes an exploratory route for the possibility of using systems of this sort to develop the understanding of the potential influence of scalar structures in astrophysical systems. We shall use, unless stated otherwise, natural units ($G=c=\hbar=1$).

\section{Framework}\label{sec:framework}
\subsection{The black hole-scalar field system}\label{sec:setup}
%
\begin{figure}%
\includegraphics[width=\columnwidth]{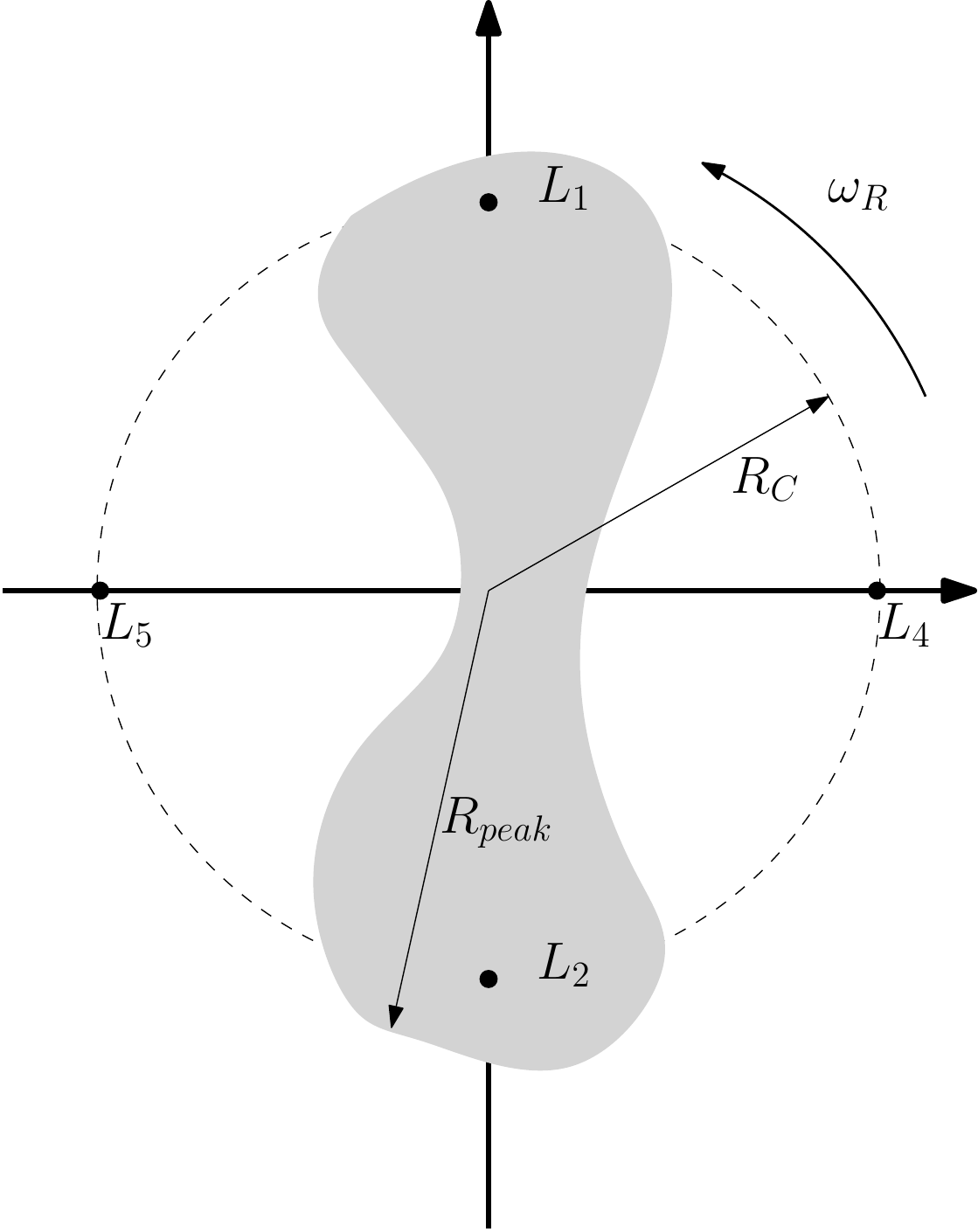}%
\caption{Schematic representation of the density $\rho$ due to the presence of the scalar field $\Phi(r,\theta,\phi)$ in the equatorial plane ($\theta = \pi/2$). The scalar field is responsible for a bar-like structure which peaks at $R_{peak}$. The points $(L_1, L_2)$ and $(L_4, L_5)$ represent the unstable and stable Lagrangian points, respectively. At these points, the radial forces acting on an orbiting particle cancel out and its motion is determined by the angular forces alone. 
Here, $R_C$ is the radial location of the Lagrangian points, known as corotation radius, and is defined in Eq.~\eqref{lagrange_resonance}. More details can be found in Appendix \ref{app:potential}.}
\label{fig:schematic-scalar}%
\end{figure}
We are interested in a massive scalar $\Phi$  minimally coupled to gravity, and described by the action
\beq
S&=&\int d^4x \sqrt{-g} \left( \frac{R}{16\pi}-\frac{1}{2}g^{\mu\nu}\Phi^{*}_{,\mu}\Phi^{}_{,\nu} -\frac{\mu^2\Phi^{*}\Phi}{2}\right)\,.\label{eq:MFaction}
\eeq
We focus on massive, non self-interacting fields. The mass $m_B$ of the boson under consideration is related to the mass parameter above through $\mu=m_{B}/\hbar$, and the theory is controlled by the dimensionless coupling
\begin{equation}
\frac{G}{c\hbar} M \mu= 7.5 \left(\frac{M }{10^8M_{\odot}}\right) \left(\frac{m_{B}c^2}{10^{-17}eV}\right)\,,\label{dimensionless_massparameter}
\end{equation}
of a massive scalar field on a curved background given by the metric $g_{ab}$.

We will focus on {\it real} scalar from now onwards~\footnote{Complex scalars admit solutions where the metric is axisymmetric~\cite{Herdeiro:2014goa,Herdeiro:2015waa}, hence most of our results do not apply to complex fields. However, most of our results do generalize should matter couple non-minimally to the scalar, for example if one considers matter which is charged under the scalar. See Sec.~\ref{sec:forces}.}. 
It can be shown that real scalars on a Kerr background admit nearly-stationary profiles. 
To a good approximation, which we will take for granted here, the scalar field is described by~\cite{Brito:2015oca,Brito:2014wla} 
\be
\Phi=A_0 g(r)\cos( \phi-\omega_R t)\sin\theta\,,\label{scalar_profile}
\ee
with
\be
g=Mr\mu^2e^{-Mr\mu^2/2}\,,
\ee
and 
\be
\omega_R \sim \mu-\frac{M^2\mu^3}{8}\,.
\ee
In principle, from the linear analysis of the problem, the amplitude $A_0$ in Eq.~\eqref{scalar_profile} can not be constrained. However, the amplitude of the scalar cloud $A_0$ is not arbitrary. We note that it should be set such that it obeys the nonlinear solutions of the Einstein-Klein-Gordon system. In general, we can express it in terms of the scalar field cloud total energy-mass $M_S$~\cite{Brito:2014wla}. For a cloud with $M_S\sim 20\% M$ and $\omega\sim \mu$, we have $A_0 \sim 0.05 (M \mu)^{2}$. We take this as our reference value.

In the limit $M\mu \ll 1$, the maximum value of the radial profile $g(r)$ is attained at $r = R_{\text{peak}}$ given by (cf. discussion around Eq. (7) in Ref.~\cite{Brito:2014wla})
\begin{equation}
\frac{R_{\text{peak}}}{M} \sim \frac{4}{(M\mu)^2} \sim 600\left(\frac{10^{-18}\,{\rm eV}}{m_Bc^2}\right)^2\left(\frac{4\times 10^6M_{\odot}}{M}\right)^2\,.
\end{equation}
This value can be used as a measure of the size of the scalar ``cloud''. 
This region is far from the BH, meaning that the curvature of spacetime is low and it is valid an analysis of the scalar field using a flat background metric~\cite{Yoshino01042014}. Accordingly, we expect that in this limit one can summarize the gravitational effects by a Newtonian gravitational potential given by a Keplerian potential

\begin{equation}
\Psi_0 = -\frac{M}{r},
\end{equation}
due to the BH, plus a small distortion
\begin{equation}
\label{eq:scalar-field-potential-schem}
\Psi_1 \sim \Psi_1^0 + \Psi_1^1 \cos(2(\phi - \omega_R t)),
\end{equation}
sourced by the scalar field in Eq.~\eqref{scalar_profile}.

The gravitational potential $\Psi_1$ is one of the effects that can be distilled after a linear analysis of Einstein's field equations (see Appendix \ref{app:weak-field}). To isolate the gravitational potential, one assumes that far from the BH, the metric is given by\footnote{Other works have employed this choice of metric - see Refs.~\cite{Khmelnitsky:2013lxt,Blas:2016ddr}.}
\begin{equation}
ds^2 = -(1 - 2\Psi_1)dt^2 + (1-2\xi)\delta_{ij} dx^i dx^j,
\end{equation}
where $\xi$ is other scalar potential which is irrelevant for non-relativistic dynamics (more details in Appendix \ref{app:weak-field}). The stress energy tensor for the scalar field is given by\footnote{Notice that the background metric is flat, as mentioned before in the text.}
\begin{equation}
T_{\mu\nu} = \frac{1}{2} \left[ \Phi^*_{,\mu} \Phi_{,\nu} + \Phi^*_{,\nu} \Phi_{,\mu} - \eta_{\mu\nu} (\eta^{\rho \sigma} \Phi^*_{,\rho} \Phi_{,\sigma} + \mu^2 |\Phi|^2) \right],
\end{equation}
and upon substitution in the Einstein's equation, the dominant contribution yields, in the non-relativistic regime of Appendix~\ref{app:weak-field},
\begin{equation}
\label{eq:poisson_scalar}
\nabla^2 \Psi_1 = - 4\pi (\rho + 3P - 3 \dot{S}),
\end{equation}
where $\rho$, $P$ and $S$ are components of the stress energy tensor defined in Appendix~\ref{app:weak-field}. We solve Eq.~\eqref{eq:poisson_scalar} in Appendix~\ref{app:potential} and we obtain that the total gravitational potential due to the presence of the BH and the scalar field is given by
\begin{equation}
\label{eq:total-potential}
\Psi = \Psi_0 + \Psi_1 = -\frac{M}{r} + \Psi_1^0 + \Psi_1^1 \cos(2(\phi - \omega_R t)),
\end{equation}
where (see Appendix \ref{app:potential} for all the details)
\begin{equation}
\Psi_1^0 = P_1(r) + P_2(r) \cos^2(\theta),
\end{equation}
and
\begin{equation}
\Psi_1^1 = P_3(r) \sin^2(\theta).
\end{equation}
In the equatorial plane, $\theta = \pi/2$, the potential is given by
\begin{align}
\label{eq:scalar_grav_potential}
\Psi = \Psi_0 + \Psi_1 = -\frac{M}{r} + P_1(r)  + P_3(r) \cos(2(\phi - \omega_R t)),
\end{align}
where $(r,\phi)$ are inertial polar coordinates in the plane and the functions $P_1(r) $ and $P_3(r)$ are defined in the aforementioned appendix. A schematic representation of the scalar field profile is shown in Fig.~\ref{fig:schematic-scalar}. We shall also see that a similar subsides when the orbiting object has a non-minimal coupling with the scalar cloud, leading, however, to different quantitative results.

The above potential has two distinctive features. First, it has a radial dependence, which can modify the structure of bound orbits of the background field, for instance changing Kepler's law\footnote{The new potential will not be proportional to $r^{-1}$ and therefore, according to Bertrand's theorem \cite{bertrand1873thorme}, not all bound orbits will be closed.}. Second, it has a periodic angular dependence on $\phi$ and $\omega$, breaking the axial symmetry of the gravitational potential. This second feature also appears in planetary motion around disks and galactic formation, and therefore can enrich the kinematics of particles around BHs. For convenience, we will organize the potential in Eq.~\eqref{eq:total-potential} specialised to the equatorial plane as
\begin{equation}
\label{eq:total_potential}
\Psi(r,\phi) = \Psi_r(r) + \delta \Psi(r,\phi),
\end{equation}
making an explicit separation between the angular and non-angular dependent components.
%
\subsection{Validity of the approximation for the orbital motion}

In the spirit of Ref.~\cite{Macedo:2013qea}, we will be using Newtonian mechanics to study the motion of a body orbiting a BH supporting a scalar field. Before doing that, we start by comparing the effective radial potential created by a Keplerian potential with the equivalent coming from a Schwarzschild BH\footnote{Although the BH that supports the scalar field is rotating, analyzing the non-rotating case is enough because in the limits we are considering the BH rotation is irrelevant.}. 
This approach is supposed to provide an approximate measure of the validity region of the Newtonian approximation.

\begin{figure}%
\includegraphics[width=\columnwidth]{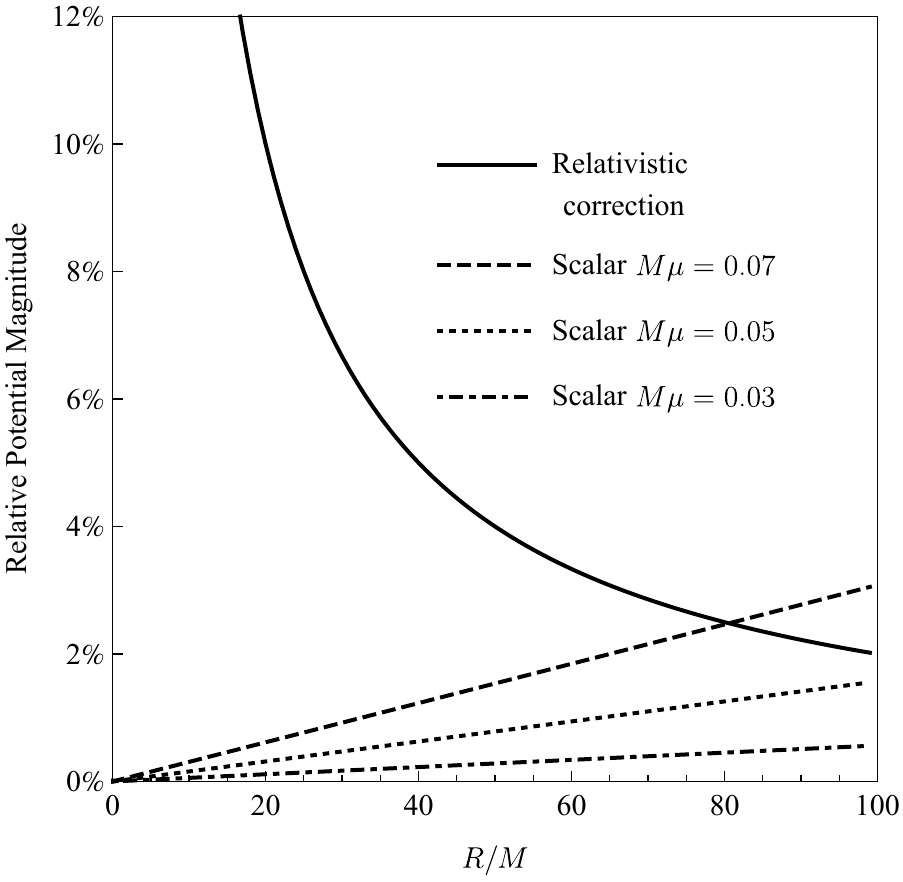}%
\caption{Relativistic and scalar field contributions to the Keplerian potential. The full line represents the ratio of the relativistic correction to the Keplerian potential, Eq.~\eqref{eq:correction} for circular orbits. The relativistic correction represents less than 10\% of the Newtonian potential for circular orbits with $R\gtrsim20 M$. The dashed, dotted and dot-dashed lines represent $|P_1|/|\Psi_0|$, the ratio of the radial part of the gravitational potential generated by the scalar field to the Keplerian potential of the BH; the value of the mass coupling $M\mu$ parameter varies, the scalar cloud comprises $20\%$ of the total mass. The influence of the gravitational potential due to the scalar field increases with distance, but overall it is very small.}
\label{fig:comp_percent}
\end{figure}

In the Newtonian approximation, the effective radial potential governing the motion (through $\dot{r}^2=E^2-V_{\text{Newton}}$) of a particle orbiting a mass $M$ is given by, 
\begin{equation}
V_{\text{Newton}} = -\frac{M}{r} + \frac{L^2}{2r^2}\,.
\end{equation}
Here $E,\,L$ are the energy and angular momentum per unit rest mass. A counterpart in the relativistic case is hard to define. We will define the analog radial potential in the Schwarzschild case by a straightforward extension using geodesics~\cite{Cardoso:2008bp}, taking dots to have the same meaning as in flat spacetime and $r$ to be the radial coordinate. One finds
\begin{equation}
V_{\text{Sch}} =  -\frac{M}{r} + \frac{L^2}{2r^2} - \frac{M L^2}{r^3}.
\end{equation}
To estimate the error of neglecting relativistic effects, we look at the magnitude of the relativistic corrections with respect to the Newtonian potential. Focusing on circular orbits of radius $R$, whose specific angular momentum is given by $L^2=M R$, the deviation from the relativistic orbit is given by
\begin{equation}
\label{eq:correction}
\frac{|\Delta V|}{V_{\rm Newton}} = \frac{|V_{\text{Sch}} - V_{\text{Newton}}|}{V_{\rm Newton}} = \frac{2M}{R},
\end{equation}
where $M$ is the mass of the central object. In Fig.~\ref{fig:comp_percent} we plot the ratio of the relativistic correction to the total energy of the classical circular orbit and from there we can see that the relativistic correction for the potential is less than $10\%$ of the total energy for orbits with radii larger than $R \sim 20 M$.
On the other hand, as we pointed out the scalar is exponentially suppressed at distances $\gtrsim 4/M \mu^2$. Thus, at very large distances, the relativistic correction dominates.
Our results will therefore be valid only for orbiting objects at distances smaller than $\sim 4/M \mu^2$.
Moreover, in order to treat the scalar field as a perturbation, the magnitude of its gravitational potential must be perturbatively smaller than the magnitude of the potential created by the central BH. In Figure \ref{fig:comp_percent}, we also plot the relative magnitude of the radial part of the scalar field potential with respect to the BH potential for different values of the mass coupling. It is sufficient to consider the radial part since the angular part is even smaller. For those particular values of mass couplings, the scalar field potential is less than $10\%$ of the BH potential. Thus, the gravitational potential due to the scalar field can indeed be treated as a perturbation to the Keplerian potential sourced by the BH.

\section{Quasi-circular orbits}\label{sec:quasicircular}

To estimate the impact of the presence of the scalar field on the dynamics of the EMRI, we will quantify the modifications it causes on circular orbits. These orbits are the simplest type of orbits in a standard EMRI and understanding how they change in response to the scalar field is a first step towards understanding how the global structure of the EMRI is modified.
\subsection{General treatment}
Under our assumptions, the study of the orbital behavior of a stellar object in the EMRI reduces to the analysis of the Lagrangian
\be
\label{eq:lag_scalar}
\mathcal{L} = \frac{1}{2} (\dot{r}^2 + r^2 (\dot{\phi} + \omega_R)^2) - \Psi(r,\phi),
\ee
which describes its motion under the influence of the potential of equation \eqref{eq:total_potential} in a system of coordinates that is corotating with the scalar field. In regions where $\frac{|P_3|}{|\Psi_r|} \ll 1$, one can obtain some insight into this system by exploring the effect of the azimuthal-dependent part on the stable circular orbits of the Keplerian potential $\Psi_0$. The perturbative approach is set up by considering the evolution of small deviations $r_1$ and $\phi_1$ to the radial and angular behavior of a stable circular orbit of radius $R_0$
\bea
r(t)& = &R_0 + r_1(t)\,,\\
\phi(t)& = &\phi_0(t) + \phi_1(t)\,,\label{eq:phi_expansion}
\eea
where $\phi_0(t) = \phi_i + (\Omega_0 -\omega_R) t$ with $\Omega_0^2 = \Psi_0'(R_0) / R_0$ and, for convenience, we fix the initial condition to be $\phi_i=0$. Neglecting second order terms in $r_1$ and $\dot{\phi}_1$, the equations of motion in the corotating frame are written as
\begin{align}
&\ddot{r}_1 + \left( \frac{\partial^2 \Psi_0}{\partial r^2} - \Omega_0^2 \right) r_1 - 2 \Omega_0 R_0 \dot{\phi}_1 + \frac{\partial P_1}{\partial r} + \frac{\partial (\delta \Psi)}{\partial r} = 0\,, \label{eq:motionr}\\
&\ddot{\phi}_1 + \frac{2 \Omega_0}{R_0} \dot{r}_1 + \frac{1}{R_0^2} \frac{\partial (\delta \Psi)}{\partial \phi} = 0\,, \label{eq:motionphi}
\end{align}
in which all derivatives are evaluated at $r=R_0$.
To study equations~\eqref{eq:motionr} and~\eqref{eq:motionphi} we consider the additional approximation $\phi_1(t) \ll \phi(t)$, i.e., $\phi(t) \sim (\Omega_0 - \omega_R)t$, meaning that we consider the perturbation $\phi_1$ so small that the angular velocity of the orbit being perturbed dominates. The equations are then written as
\begin{align}
\ddot{r}_1 &+ \left( \frac{\partial^2 \Psi_0}{\partial r^2} - \Omega_0^2 \right) r_1 - 2 \Omega_0 R_0 \dot{\phi}_1 \nonumber\\
& + \frac{\partial P_1}{\partial r}  + \frac{\partial P_3}{\partial r} \cos(2 (\Omega_0 - \omega_R) t) = 0 \,,\label{eq:motionr2}\\
\ddot{\phi}_1 &+ \frac{2 \Omega_0}{R_0} \dot{r}_1 -  \frac{2}{R_0^2} P_3 \sin(2 (\Omega_0 - \omega_R) t) = 0\,,  \label{eq:motionphi2}
\end{align}
where the coefficients are evaluated at $r = R_0$. Integrating Eq.~\eqref{eq:motionphi2} and substituting the result in \eqref{eq:motionr2} one obtains
\begin{align}
  \ddot{r}_1 + & \left(\frac{\partial^2 \Psi_0}{\partial r^2} + 3\Omega_0^2 \right) r_1 + \frac{\partial P_1}{\partial r}\nn\\
                 &= - B(R_0) \cos(2(\Omega_0 - \omega_R)t),
\label{eq:eq_ddr}
\end{align}
whose general solution is given by
\begin{align}
r_1(t) &= A \cos(\kappa_0 t + \alpha) - B(R_0) \frac{\cos(2 (\Omega_0 - \omega_R) t)}{\kappa_0^2 - 4(\Omega_0 - \omega_R)^2} \nonumber\\
&- \frac{C(R_0)}{\kappa_0^2}\label{eq:r1_sol},\\
  \phi_1(t) &= -\frac{2 \Omega_0 A}{R_0 \kappa_0} \sin(\kappa_0 t + \alpha) + D(R_0) \sin(2 (\Omega_0 - \omega_R) t)\nn\\
  &- \frac{C(R_0)}{\kappa_0^2}t\label{eq:phi1_sol},
\end{align}
with $\kappa^2_0 = \Psi''_0 + 3 \Omega_0^2$ and 
\begin{align}
&A = \left[r_{1i} - \frac{B}{\kappa_0^2 - 4(\Omega_0 - \omega_R)^2} + \frac{C}{k_0^2}\right]\cos^{-1}\alpha\label{eq:amplitude-general},\\
&\tan\alpha = \dot{r}_{1i}^{-1}\kappa_0 \left[ \frac{C}{k_0^2} -r_{1i} - \frac{B}{\kappa_0^2 - 4(\Omega_0 - \omega_R)^2}\right]\label{eq:phase-general},\\
&B= \frac{\partial P_3}{\partial r}  + \frac{4 \Omega_0 P_3}{R_0 (\Omega_0 - \omega_R)},\\
&C=  \frac{\partial P_1}{\partial r}\label{eq:CR0},\\
&D= \frac{ \Omega_0 B}{R_0(\kappa_0^2 - 4(\Omega_0 - \omega_R)^2)(\Omega_0 - \omega_R)}\nn\\
       & - \frac{P_3}{R_0^2 (\Omega_0 - \omega_R)^2}\label{eq:DR0},
\end{align}
where all the quantities are calculated at $R_0$ and $(r_{1i},\dot{r}_{1i})$ are the initial conditions for the radial motion.
This kind of solution is long known in problems with non-axisymmetric potentials (see, e.g., Refs.~\cite{Berman1977ApJ...216..257B,Berman1979ApJ...231..388B,Contopoulos1973ApJ...181..657C,Contopoulos1978A&A....64..323C,binney2011galactic}). We can readily see the presence of some singularities in Eqs.~\eqref{eq:r1_sol}, \eqref{eq:amplitude-general} and \eqref{eq:DR0}. Two of the singularities appear when
\begin{equation}
  \kappa_0 = \pm 2(\Omega_0 - \omega_R)\,.
\end{equation}
These are called Lindblad (inner and outer) resonances. The other singularity, given by
\begin{equation}
  \Omega_0 = \omega_R,
\end{equation}
is called co-rotating resonance, because the perturbation is being made to a circular orbit which is synchronized with the potential (in this case with the scalar cloud). When a resonant frequency is approached, the above linear analysis breaks down. We shall look into these particular orbits in the following sections.
The radii at which the outer (inner) Lindblack resonance occurs will be termed outer (inner) Lindblad radius $R_{L\pm}$. The radius at which the co-rotating resonance occurs is the co-rotation radius $R_C$.
\begin{figure}%
\includegraphics[width=\columnwidth]{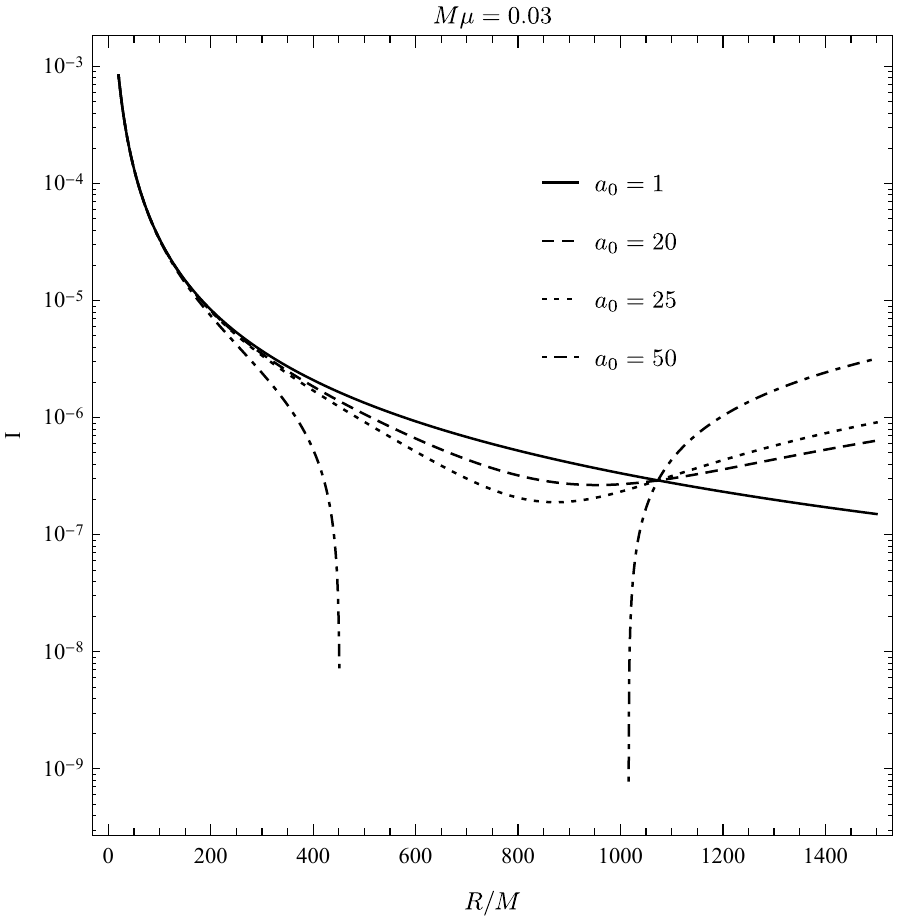}%
\caption{Representing the instability measure $I$ of Eq.~\eqref{eq:instability-quantity} as a function of the radius of the circular orbit for mass coupling $M\mu=0.03$ 
and different scalar-field amplitudes. Large scalar amplitudes give rise to a set of unstable orbits. Notice that the range of radii in which the instability measure is negative, depends on the value of the parameter $a_0$.}
\label{fig:instability_plot}
\end{figure}
%
\subsubsection{Circular orbits}\label{subsubsec:circular orbits}
Aside from giving rise to resonant orbits, the angular part of the scalar field potential $\Psi_1$ plays no other role in the dynamics of the system due to its small value (recall that the amplitude of the angular part of the potential is smaller than the radial part). Indeed, the other effects of the presence of the scalar field are an exclusive consequence of the radial part of its gravitational potential. 
To further explore its effects, we ignore the presence of the $\phi$-dependent part of the potential $\Psi_1$ in the Eqs.~\eqref{eq:motionr2} and \eqref{eq:motionphi2}, which means that the angular momentum is exactly conserved, and the solution in given by Eqs.~\eqref{eq:r1_sol} to \eqref{eq:phase-general} with  $B(r) \equiv 0$ and $D(r) \equiv 0$. 
As in the analysis with the angular part of the potential, the behavior of the perturbations indicates that the circular stable orbits of the BHSFS are not exactly Kleperian. 
We start with initial conditions at $r_{1i}=\dot{r}_{1i}=0$, and initial radius such that the orbit would be circular -- for the same central mass -- if the scalar cloud did not exist. We find a solution that deviates from the Keplerian circular orbit. This, of course, is expected: given a value of the angular momentum, the corresponding value of the radius of the circular orbit of the total radial potential $\Psi_r$ is different from the radius of the circular orbits of $\Psi_0$. Quantifying this radial difference is a way of looking into the influence of the scalar field on the orbital structure around the SMBH. We will indicate the radii of the circular orbits of the total potential $\Psi_r$ by $R_0^*$; their values are given, for fixed angular momentum $L$, by
\begin{equation}
  \frac{L^2}{(R_0^*)^3} = \frac{d \Psi_r}{dr},
  \label{eq:scalar-circ-radius}
\end{equation}
where the derivative is taken at $R_0^*$. The stability of these orbits is guaranteed as long as
\begin{equation}
  \label{eq:instability-quantity}
I\equiv   \frac{d \Psi_r}{dr} + \frac{R_0^*}{3} \frac{d^2\Psi_r}{dr^2} > 0.
\end{equation}
This inequality is always verified in the range of the mass coupling parameter we are considering, thus all circular orbits of the potential $\Psi_r$ are stable. Notice, however, that the amplitude of the scalar field can be larger---for instance in non-minimal coupling scenarios. Parametrizing the amplitude of the scalar field as
\begin{equation}
  \label{eq:param-amp}
  \tilde{A}_0 = a_0 A_0,
\end{equation}
where $A_0$ is the standard amplitude of the scalar field (see Sec.~\ref{sec:setup}), we observe that for $a_0>a_0^{\rm insta}$, unstable orbits appear. More precisely, for $a_0 < a_0^{\rm insta}$ all circular orbits are stable and for $a_0 > a_0^{\rm insta}$ a window of unstable circular orbits with $R_{\rm min}<R_0^{*}<R_{\rm max}$ exists. This is shown in Fig.~\ref{fig:instability_plot}, where it is represented the quantity described in Eq.~\eqref{eq:instability-quantity}. Furthermore, it is also observed that after crossing the boundary imposed by $a_0^{\rm insta}$, the range $\{R_{\rm min},R_{\rm max}\}$ increases with $a_0$.

\begin{figure}%
\includegraphics[width=\columnwidth]{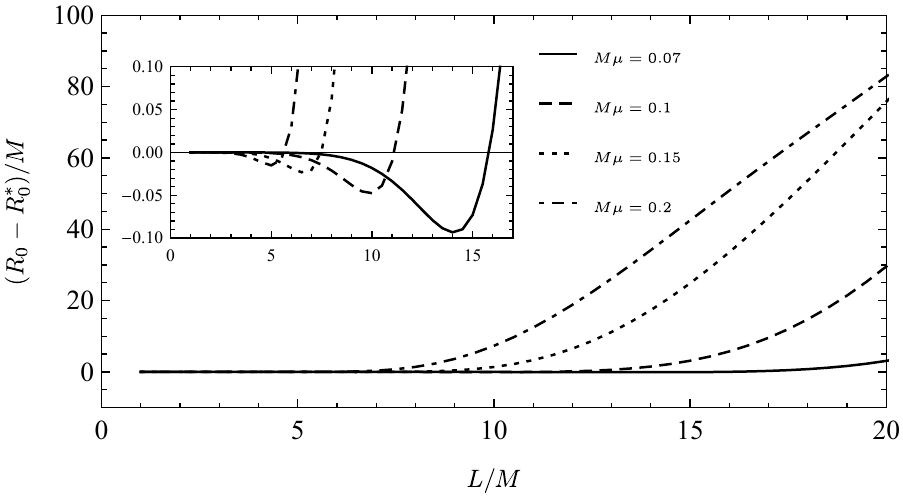}%
\caption{Difference between the radius of a Keplerian circular orbit with angular momentum $L$ and the radius of a circular orbit of the potential $\Psi_r$ (Eq.~\eqref{eq:total_potential} ) with the same angular momentum. For small values of the angular momentum, the difference is negligible, becoming negative up to the value $L = \sqrt{R_{\rm peak}}$ (see Eq.~\eqref{eq:radius-inver}), where $R_0 = R_0^*$; after this value the difference is positive and grows indefinitely. This growth means that, far from the SMBH, a stable circular orbit of the BHSF system with a given angular momentum has a smaller radius than its Keplerian counterpart (in which the Kepler potential is generated only by the SMBH).}%
\label{fig:radii_dif}%
\end{figure}

\begin{figure}%
\includegraphics[width=\columnwidth]{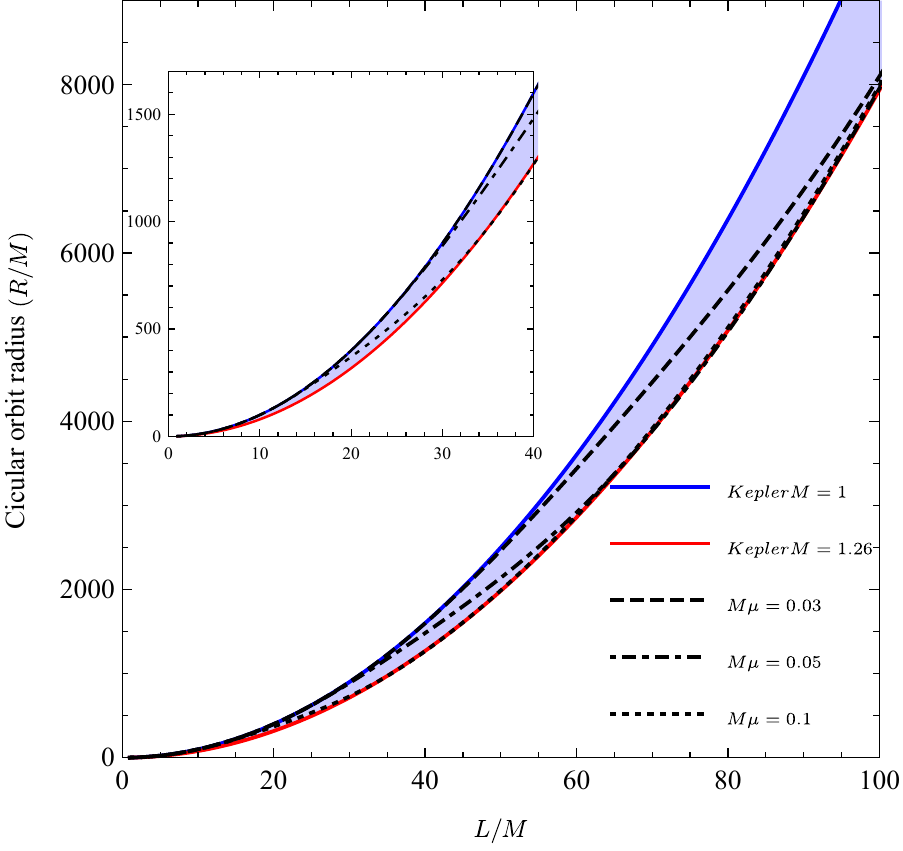}%
\caption{Representing the value of the radii of stable circular orbits as a function of the angular momentum. Continuous lines refer to circular orbits of a Keplerian potential $-M/r$ -- the blue line corresponds to a Keplerian potential with $M=1$ and the red line to $M=1.26$. Discontinuous lines represent the radii of circular orbits for the potential of the BHSF system $\Psi_r$; for small values of the angular momentum, these values are similar to those generated by a Keplerian potential with $M=1$ while for large values of the angular momentum they stabilize to the curve described by the Keplerian potential with $M = 1.26$. This fact leads us to the conjecture that any other mass coupling parameter would generate a plot that would be bounded by the two Keplerian curves; the only influence of the mass coupling parameter is the extent to which the radii of circular orbits deviate from a Keplerian relation, as can be seen in the inline. In fact, the bigger the mass coupling parameter the smaller is the range of the deviation.}%
\label{fig:radial_parabola}%
\end{figure}

For small values of the angular momentum and mass coupling parameter, the difference between the radius of Keplerian circular orbits, $R_0$, and the radius of circular orbits of the BHSFS, $R_0^*$, is negligible\footnote{We keep the angular momentum fixed. This observation shows that close to the SMBH Keplerian circular orbits are good approximations for the circular orbits of the whole system. This reinforces the validity of the results we obtained for the location of the resonant orbits.}; however, this difference has a non-trivial evolution once we start to vary those parameters; this can be appreciated in Fig.~\ref{fig:radii_dif}. As the angular momentum increases, the difference between the radii has a behavior which is controlled by the value of the mass coupling parameter: large values of $M\mu$ imply a large radii difference for a fixed value of the angular momentum. Notwithstanding, the difference in radii at large distances is due to the fact that now the orbiting particle sees a different effective mass (central object plus the scalar field surrounding it, see Fig.~\ref{fig:radial_parabola}). In Fig.~\ref{fig:radii_dif} we also see that, for each value of the mass coupling parameter, there is a value of the angular momentum for which the radii of the stable circular orbits is equal for both potentials; the corresponding value was numerically determined to be
\begin{equation}
  R_0 = R_0^* = R_{\rm peak} \sim \frac{4M}{(M\mu)^2}.
  \label{eq:radius-inver}
\end{equation}
This value represents the location where the scalar cloud typically peaks \cite{Brito:2014wla}.
Another way of looking at the difference between the circular orbits of an isolated BH and a scalar-surrounded one is to observe that close to the SMBH a potential $\Psi = -M/r$ is dominant, while far from the SMBH the dynamics are dominated by $\Psi = -M_{\rm eff}/r$ in which $M_{\rm eff}$ is an effective mass value, in units of the mass $M$ of the SMBH. It was numerically found that our system has $M_{\rm eff} \sim 1.26$, as can be seen in Fig.~\ref{fig:radial_parabola}. This number can be interpreted by looking at Eq.~\eqref{eq:scalar-circ-radius} in the form
\begin{equation}
  L^2 = R_0^* + {R_0^*}^3 C(R_0^*),
\end{equation}
and observing that for large values of angular momentum $L$ and radius $R_0^*$ it can be written as
\begin{equation}
  L^2 = R_0^* + {R_0^*}^3 \left(\frac{32\pi a_0^2}{{R_0^*}^2} 	\right),
\end{equation}
where the amplitude of the scalar field $A_0$ was again written as $A_0 = a_0 (M\mu)^2$. Using the value for $a_0$ prescribed in Sec.~\ref{sec:setup} we obtain
\begin{equation}
  L^2 = 1.25 R_0^*
\end{equation}
which is close to the value obtained numerically\footnote{Recall that the circular orbit of a Keplerian potential $-M/r$ with angular momentum $L$ has a radius given by $R = L^2/M$.}. We see, then, that far from the SMBH the potential governing the dynamics is still Keplerian, but the mass sourcing it is not the SMBH mass. This  ``effective" mass $M_{\rm eff}$ corresponds to the mass of the SMBH plus the total mass contained the scalar field.
%
\subsubsection{Rate of precession}
%
\begin{figure}%
\includegraphics[width=\columnwidth]{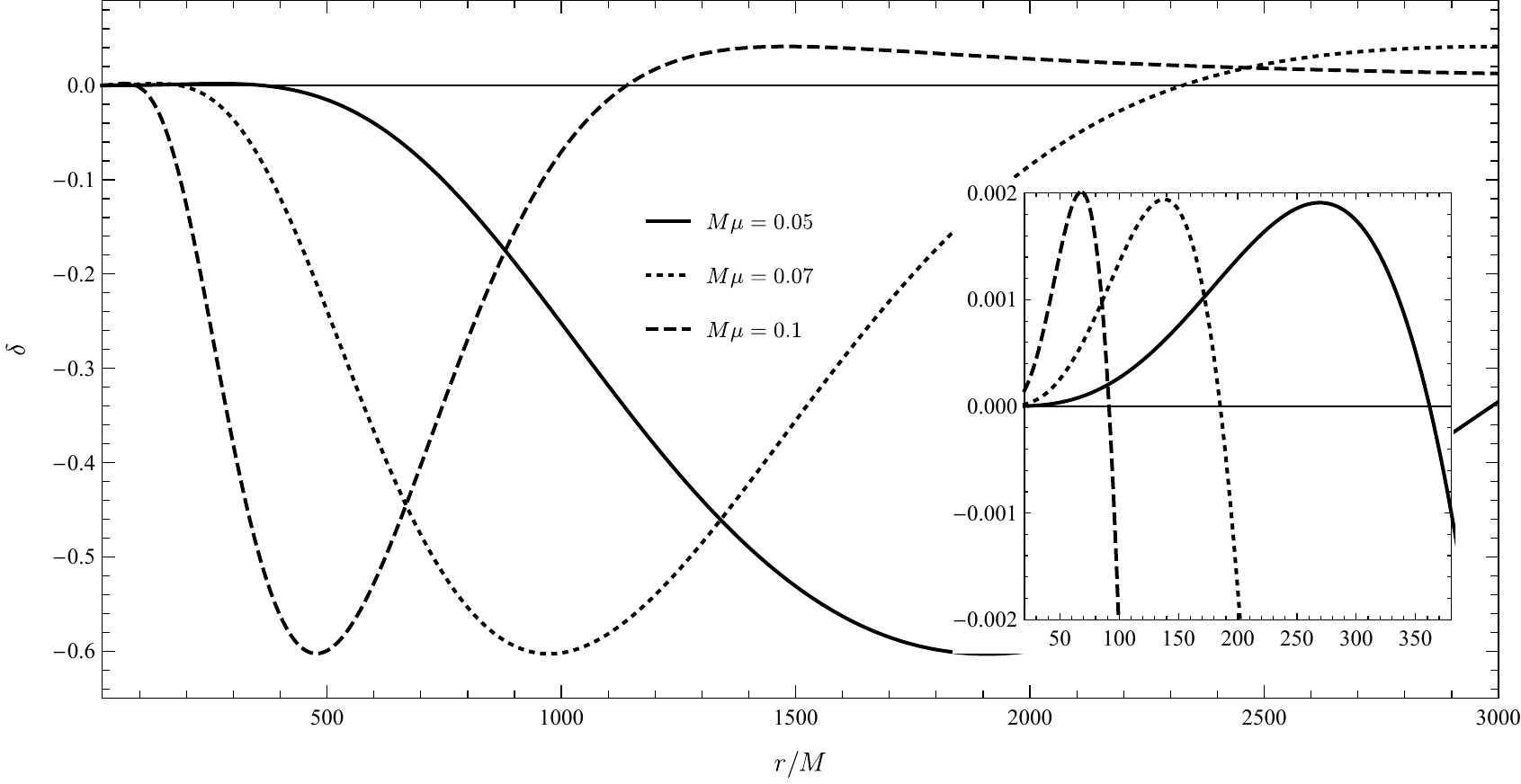}%
\caption{Precession rate as a function of distance from the SMBH, for different mass coupling parameters $M\mu$. The maximum and mininum precession-rate values are -- to a good precision -- independent of the mass coupling parameter. The precession rate takes positive and negative values at different distances from the SMBH depending on the value of the mass coupling parameter.
The zoom in shows that there are, in fact, two zeroes of the precession rate, see Eq.~\ref{zeroes}.}
\label{fig:apsidal}
\end{figure}
Another expected feature of the potential generated by the presence of the scalar field is the precession of the orbits. The apsidal angle of a quasi-circular orbit obtained from a perturbation of a circular orbit of radius $R_0^*$ is
\begin{equation}
  \psi = \pi \left[ 3 + R_0^* \frac{\Psi_r''(R_0^*)}{\Psi'(R_0^*)} \right]^{-1/2}.
\end{equation}
From this expression, we obtain that in each cycle around the SMBH, the orbit precesses at a rate $\delta = 2(\psi - \pi)$ radians. In this particular case, 
\begin{equation}
\delta = 2\pi \left(\sqrt{\frac{1 + (R_0^*)^2 C(R_0^*)}{2 + 3(R_0^*)^2 C(R_0^*) + (R_0^*)^3 C'(R_0^*)}} -1 \right)
\end{equation}
where $C(r)$ is given by Eq.~\eqref{eq:CR0}. The dependence of this precession rate on the mass coupling parameter and on the radius of the corresponding orbit is represented in Fig.~\ref{fig:apsidal}. Two features are instantly noticed: a) the precession rate changes sign, indicating that orbits can precess both clock- and counter-clock wise, and b) there is a maximum and a minimum value of the precession rate that is independent of the mass coupling parameter of the system, indicating that it must be an intrinsic characteristic of the model -- it is the distance from the BH at which these extrema occur that depend on the mass coupling parameter. 
The precession rate changes sign twice: the first time being from a region of positive precession rate to a region of negative precession rate (top-right panel of Fig.~\ref{fig:apsidal}), the second time being from a region where the precession rate is negative to a region in which the precession rate is positive (bottom panel of Fig.~\ref{fig:apsidal}). 
As one moves farther from the SMBH the precession rate stabilizes to zero, meaning that a Keplerian potential is dominating the dynamics\footnote{This is confirmed by the results of Fig.~\ref{fig:radial_parabola}.}. The existence of these localized regions of positive and negative precession rate, each of them with well defined maximum and minimum values offer a good source of phenomenology that may allow for a characterization of the system. We found numerically that both zeros are related with the mass coupling parameter as
\be
M\mu^2R_{\rm zeroes}= (1,\,11)\,.\label{zeroes}
\ee
%
\subsection{Resonant orbits}\label{sec:quasicircular_resonances}
%
\begin{figure}%
\includegraphics[width=\columnwidth]{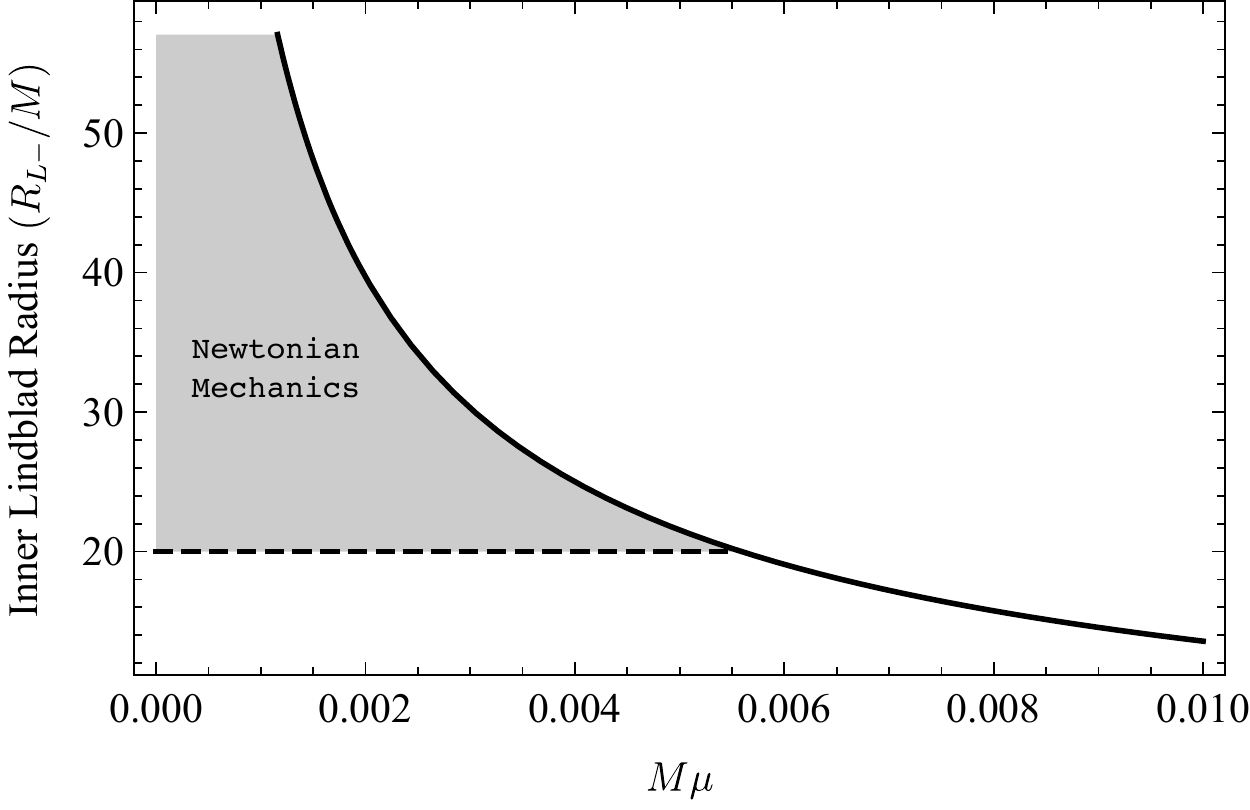}%
\caption{
Radius of the innermost Lindblad radius, $R_{L-}$, as function of the coupling parameter $M\mu$. The grey area highlights the region in which Newtonian approximation is suitable (see Fig.~\ref{fig:comp_percent}), which is given by $M \mu \lesssim 0.006$.
}%
\label{fig:inner_Lindblad_Newtonian_Limit}%
\end{figure}
For the rest of this section, we will go back to Eqs.~\eqref{eq:r1_sol} and \eqref{eq:phi1_sol}, focusing on the resonant orbits. For a particle near the resonant orbits, the general perturbative approach presented in those equations is not adequate since it gives unphysical behavior for the perturbations\footnote{The complete understanding of this problem is out of the scope of our work. More details can be found in Refs.~\cite{arnol1963small,henrard2005adiabatic}.}. Our main motivation to pursue a more detailed analysis of these orbits is their important role in the so called angular momentum transfer mechanisms in the context of galactic dynamics~\cite{Goldreich:1980wa,1979ApJ...233..857G,1972MNRAS.157....1L}; we suspect that most---if not all---of those results can be generalized to the context of BHs and scalar fields (bearing in mind, naturally, the different lengthscales). In other words, it is possible that BHs anchoring scalar fields may give rise to galactic-like structure on lengthscales of a few hundred Schwarzschild radii.

The radii of the circular orbits that correspond to the resonant frequencies are obtained by substituting the expressions for the angular frequency $\Omega(r)$ and the epicyclic frequency $\kappa(r)$ in the equations defining the resonances and solving for the radial coordinate. By doing this, one can immediately see that the smallest of the three resonant radii is the one that corresponds to the inner Lindblad frequency. In order to guarantee that Newtonian mechanics can be used to study the inner Lindblad resonant orbit, its radius should be $R_{L-}\gtrsim 20 M$, see Fig.~\ref{fig:comp_percent}. This scale can be controlled by the mass coupling $M \mu$ given that the scalar field rotates with angular frequency $\omega_R \sim \mu$; taking this into account, the inner Lindblad radius is given by
\begin{equation}
\frac{R_{L-}}{M} \approx \l({\frac{1}{4M^2\mu^2}}\r)^{1/3},
\end{equation}
from which one can estimate the maximum value of $M \mu$ such that a Newtonian analysis is justified. This is shown in Fig.~\ref{fig:inner_Lindblad_Newtonian_Limit}. 
Once the inner Lindblad radius is sufficiently far from the central BH such that Newtonian mechanics is valid, then the other two resonances (corotation and outer Lindblad resonance) are automatically ensured to be within the same regime. The corotation and outer Lindblad radii are, respectively,
\begin{equation}
\frac{R_C}{M} \approx \l({\frac{1}{M^2\mu^2}}\r)^{1/3}, \quad \frac{R_{L+}}{M} \approx \l({\frac{9}{4M^2\mu^2}}\r)^{1/3},\label{lagrange_resonance}
\end{equation}
i.e., $R_{L-} < R_C < R_{L+}$. 

The analytical solutions we shall be presenting for the quasi circular resonant orbits follow from the same assumptions made for the general quasi-circular orbits, i.e., the perturbations $r_1$ and $\phi_1$ will be considered small. To do this, we take the equations of motion for the perturbations and analyze them separately for each of the three resonant frequencies mentioned previously.

\subsubsection{Lindblad Resonances}
To study the behavior of the system at the Lindblad resonances, we have to go back to equations \eqref{eq:motionr2} and \eqref{eq:motionphi2} and make the explicit substitution
\be
R_0 \rightarrow R_{L\pm} \quad \Omega_0 \rightarrow \omega_R \pm \frac{1}{2} \kappa_{L\pm},
\ee
which under the same reasoning applied before will allow us to write
\be
\ddot{r}_1 + \kappa_{L\pm}^2 r_1 + C(R_{L\pm}) + \tilde{B}(R_{L\pm}) \cos(\kappa_{L\pm}t) = 0,
\ee
with
\begin{equation}
  \tilde{B}(R_{L\pm}) = \frac{\partial P_3}{\partial r}  \pm \frac{4 \Omega_0 P_3}{R_{L\pm} \kappa_{L\pm}}.
\end{equation}
The differences between the equations for the inner and the outer Lindblad orbits are the numerical value of the epicyclic frequency, the sign in the equation of motion for $\phi_1$ (see Eq.~\eqref{eq:phi1Lindblad}) and the functions $B(r), C(r)$. The previous equation has a direct analytic solution given by
\begin{align}
  \label{eq:r1Lindblad}
  r_1(t) &= - \frac{1}{\kappa_{L\pm}^2} \bigg[ 2 C(R_{L\pm}) + (\tilde{B}(R_{L\pm}) - 2 \kappa_{L\pm}^2 \Gamma_1) \cos(\kappa_{L\pm}t)\nn\\
  & + \kappa_{L\pm}^2(\tilde{B}(R_{L\pm}) t - 2\kappa_{L\pm}\Gamma_2) \sin(\kappa_{L\pm}t) \bigg],
\end{align}
where $\Gamma_1$ and $\Gamma_2$ depend on the initial conditions as
\begin{align}
  \Gamma_1& = \frac{1}{2 \kappa_{L\pm}^2} \left[ r_{1i} \kappa_{L\pm}^2 + 2 C(R_{L\pm}) + \tilde{B}(R_{L\pm})\right]\label{eq:LindConst1}\\
  \Gamma_2& = \frac{1}{2 \kappa_{L\pm}^2} \dot{r}_{1i}\label{eq:LindConst2}.
\end{align}
Using this expression for the evolution of $r_1$ in
\begin{equation}
  \label{eq:phi1Lindblad}
\dot{\phi}_1 + \frac{2(\omega_R \pm \frac{1}{2}\kappa_{L\pm})}{R_{L\pm}} r_1 \pm \frac{2 P_3(R_{L\pm})}{\kappa_{L\pm} R_{L\pm}^2} \cos(\kappa_{L\pm} t) = 0,
\end{equation}
one can derive the expression for $\phi_1$.

\begin{figure}%
\includegraphics[width=0.85\columnwidth]{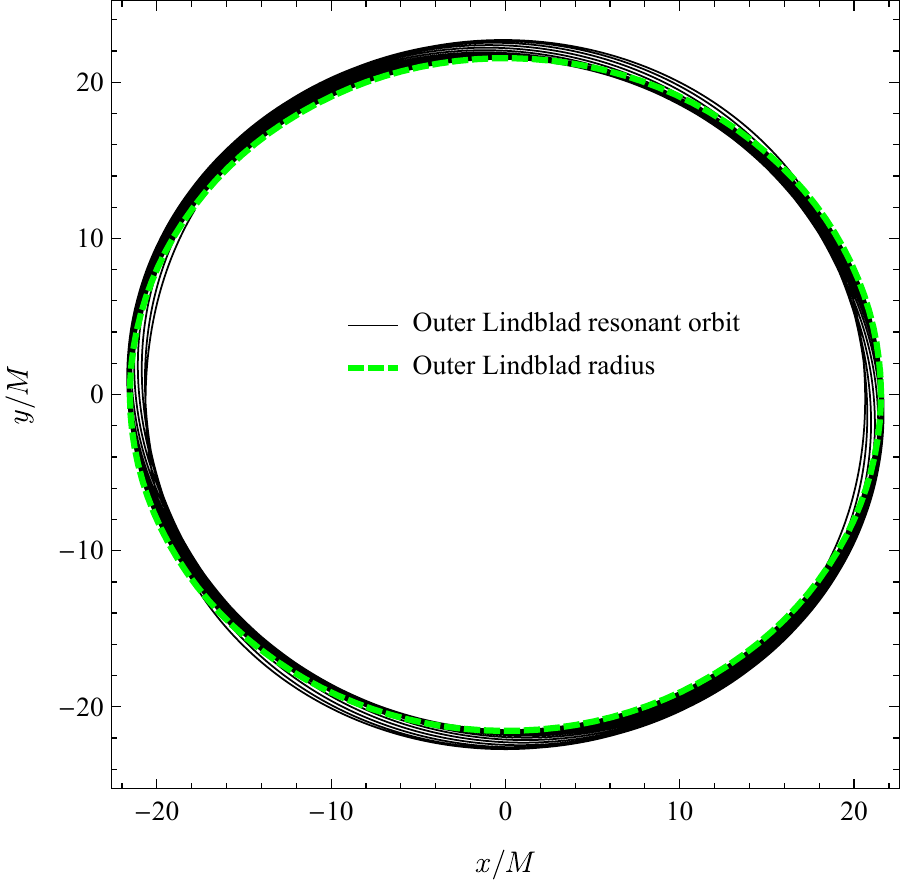}%
\caption{Representing the orbit of a particle at the outer Lindblad resonance. We used $M\mu = 0.015$ and we artificially enhanced the scalar field amplitude (again, we artificially increased the amplitude to $\tilde{A}_0 = 350 A_0$, where $A_0$ is the standard amplitude of the scalar field given by $A_0 = 0.05 (M\mu)^2$) to allow for an easier representation of the main characteristics. The behavior presented is described by a first order approximation (see Eq.~\eqref{eq:r1Lindblad}) where an increase in the radius of the orbit can be seen.
}
\label{fig:lindblad-orbit}
\end{figure}

The solutions for $r_1$ and $\phi_1$ at Lindblad resonances show a different behavior from the general case \eqref{eq:r1_sol} and \eqref{eq:phi1_sol}, with the radial perturbation growing significantly even when the orbiting body is initially placed in a circular orbit, i.e., when $r_{1i} = \dot{r}_{1i} = 0$. Note that we chose initial conditions to correspond to a circular orbit {\it in the absence of a scalar cloud}. Because of the term proportional to the time parameter in Eq.~\eqref{eq:r1Lindblad}, the radial perturbation increases at each period of oscillation -- see Fig.~\ref{fig:lindblad-orbit}. This behavior results from the fact that up to first order, the perturbation $r_1$ is described by the equation of motion for a harmonic oscillator (with natural frequency given by $\kappa_{L\pm}$) being excited by a harmonic force with the same frequency. This is a classical example of resonance.
\begin{figure}%
\includegraphics[width=0.8\columnwidth]{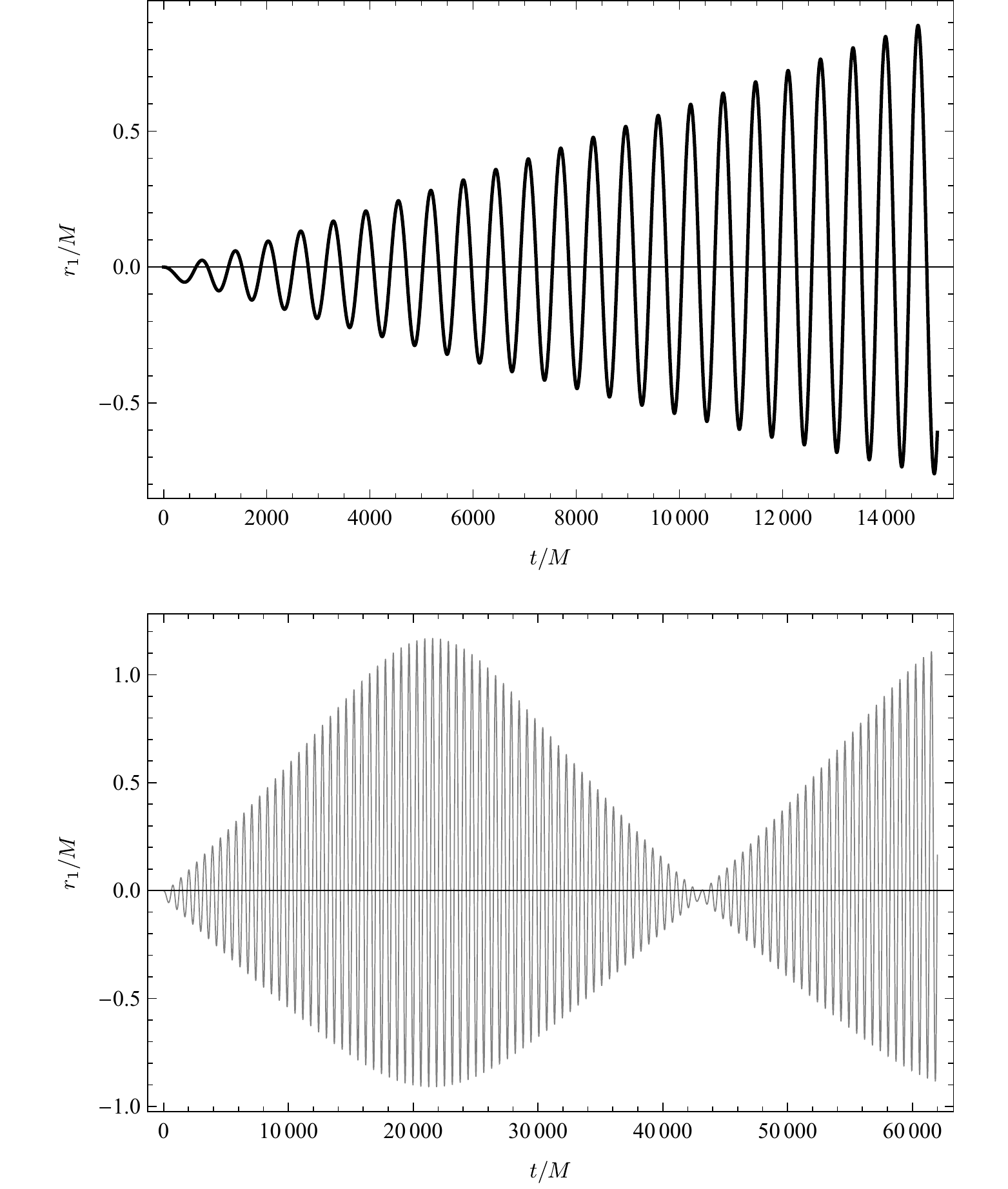}%
\caption{
 Representing $r_1(t) = r(t) - R_{L+}$. Notice that this calculation was made using the initial conditions $\dot{r}_{1i} = r_{1i} = 0$, meaning that the particle was initiated in the circular orbit with radius given by the outer Lindblad radius. The amplitude of the scalar field is artificially enhanced (we use an amplitude $\tilde{A}_0 = 350 A_0$ where $A_0$ is the standard amplitude of the scalar field given by $A_0 = 0.05 (M\mu)^2$) in order for the beatting pattern to be more easily observed. The mass coupling parameter used is $M\mu = 0.015$ so that the outer Lindblad radius is $R_{L+} \sim 21.5 M$. \textit{Top panel:} The first order evolution of the perturbation $r_1$ is represented, agreeing with the analytical expression of Eq.~\eqref{eq:r1Lindblad} (see Fig.~\ref{fig:lindblad-orbit} for a depiction of the orbit). \textit{Bottom panel:} As the absolute value of the perturbation increases due to the first order resonant behavior, higher orders of the equation of motion acquire importance preventing an indefinite grow of $r_1$ by giving rise to a beat pattern.
}%
\label{fig:beat_pattern}%
\end{figure}
While the first order approximation holds, the value of the radial perturbation increases as seen in Fig.~\ref{fig:beat_pattern}. Once this approximation stops being valid, which eventually happens if enough time passes, the higher order components of the equations of motion describing $r_1$ and $\phi_1$ become important and the evolution of the radial perturbation is no longer described by Eq.~\eqref{eq:r1Lindblad}; the higher order terms force $r_1$ to decrease, as can be seen in Fig.~\ref{fig:beat_pattern}, ending up describing a beating pattern. The time that it takes for a complete beat, both at inner and outer Lindblad resonances, depends on the mass coupling parameter and on the amplitude $a_0$ of Eq.~\eqref{eq:param-amp}; we numerically compute this dependence, finding a good fit to be
\begin{equation}
  \tau_{\text{beat}} \sim \frac{1}{a_0^{1.3}} \frac{21.5 M}{(M\mu)^{\frac{91}{25}}} \,,
\end{equation}
which means that the analytic solutions of Eq.~\eqref{eq:r1Lindblad} and \eqref{eq:phi1Lindblad} are accurate up to $\tau_{\text{beat}} / 2$ in which the maximum value of $r_1$ is attained.
%
%
\subsubsection{Corotation Resonance}
\begin{figure}[!tbp]
\includegraphics[width=.7\columnwidth]{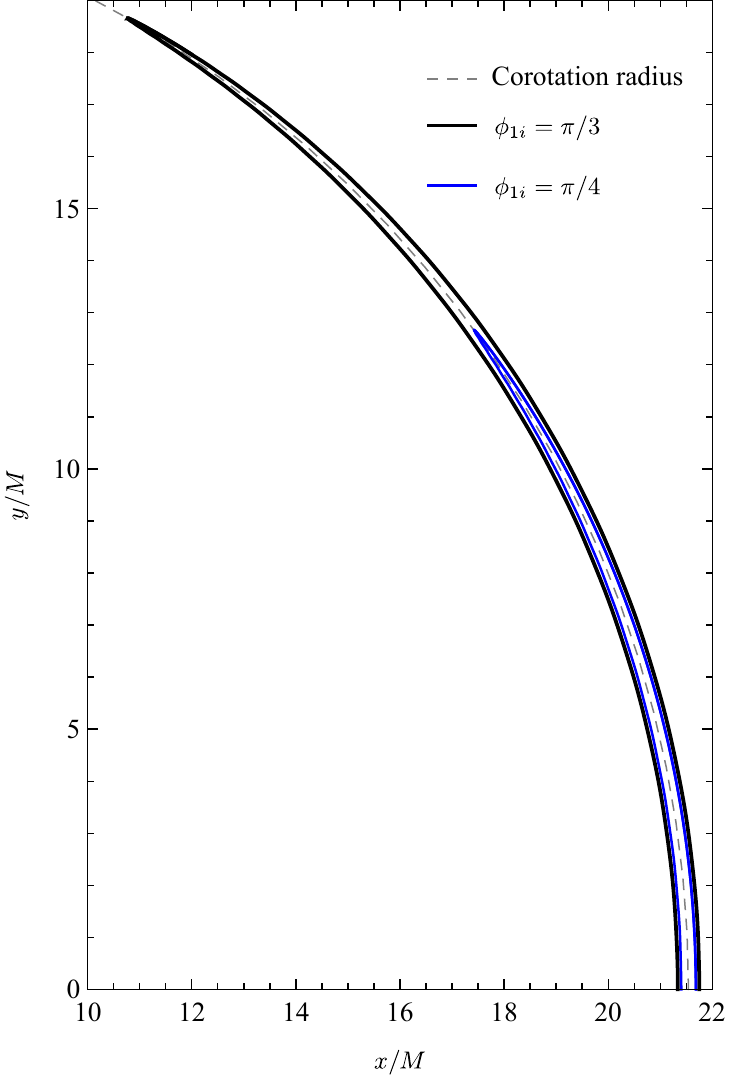}
\caption{Representing banana orbits. The mass coupling parameter used in the calculations is $M\mu = 0.01$ so that the corotation radius is given by $R_C \sim 21.5 M$. The orbits are depicted in the co-rotating frame and because they are symmetric in $y$ we show only one quadrant. We used an artificially enhanced amplitude for the scalar field (we use $\tilde{A}_0 = 350 A_0$) in order to make possible a clearer representation of the orbits. We show two initial angles, $\phi_{1i}=\pi/3$ and $\phi_{1i}=\pi/4$. The influence of the initial angle on the extent and on the width of the banana orbit is apparent, and can be related to the approximate analytic solution presented in Table~\ref{tab:initial-conditions}.}
\label{fig:compare_corotation}
\end{figure}

Going back to the equation of motion~\eqref{eq:motionr} and~\eqref{eq:motionphi}, considering that the circular orbit has a radius given by the corotation radius we see that the zeroth order term vanishes identically and (see equation \eqref{eq:phi_expansion})
\be
\phi(t) = \phi_i + \phi_1(t).
\ee
Thus, what sets corotation apart from the general orbits and Lindblad orbits is the fact that besides the initial condition of the radial perturbation, also the initial angle is important for the motion, as can be seen in the corotation equations of motion
\begin{align}
\ddot{r}_1 &+ \left( \frac{\partial^2 \Psi_0}{\partial r^2} - \omega_R^2 \right) r_1 - 2 \omega_R R_C \dot{\phi}_1 + C(R_C) \nn\\
&  + \frac{\partial P_3}{\partial r}\left[\cos(2\phi_i) - 2\sin(2\phi_i)\phi_1\right] = 0 \,,  \label{eq:eom_r1_corot}\\
\ddot{\phi}_1 &+ \frac{2 \omega_R}{R_C} \dot{r}_1 -  \frac{2  P_3}{R_C^2}\left[\sin(2\phi_i)+2\cos(2\phi_i)\phi_1\right] = 0,  \label{eq:eom_phi1_corot}
\end{align}
where the coefficients are computed at the corotation radius. To write these equations, we considered that the angular perturbation $\phi_1$ is small enough to allow for the expansion of the corresponding sinusoidal functions up to first order. Moreover, we observed that in order for the analytical solution to be valid, the initial condition $\phi_{1i} = \phi_1(t=0)$ has to be of the same order as $\phi_1$\footnote{The approach presented here focuses on motions around the point $\phi_i = 0$, a stable Lagrangian point. For an alternative approach to the study of the motion at corotation, see Section 3.3 of Ref.~\cite{binney2011galactic}.}. After all these cautionary remarks, one can advance to the solution of the equation of motion; the method used previously to obtain the expressions describing the motion of the orbiting body is not adequate in this case. One has to employ a more evolved method, described in Appendix \ref{app:corotation}, which in general gives a solution of the form
\begin{align}
r_1(t) = C_1 \cos(\omega_1 t) + C_2 \cos(\omega_2 t)\label{eq:r1corot},\\
\phi_1(t) = C_3 \sin(\omega_1 t) + C_4 \sin(\omega_2 t)\label{eq:phi1corot},
\end{align}
or
\begin{align}
r_1(t) = C_1 \sin(\omega_1 t) + C_2 \sin(\omega_2 t)\label{eq:r1corot1},\\
\phi_1(t) = C_3 \cos(\omega_1 t) + C_4 \cos(\omega_2 t)\label{eq:phi1corot1},
\end{align}
where $\omega_i$ depends only on the parameters of the problem---both the scalar field and the mass of the SMBH---and $C_i$ depends on the parameters of the problem and the initial conditions. Comparison with the numerical calculations (see Table~\ref{tab:initial-conditions}) shows good agreement with the analytical solutions for an arbitrary range of the time parameter. In general, one of the sinusoidal functions in the solution for $r_1$ and $\phi_1$ dominates over the other leading to the so called \textit{banana orbits} \cite{Contopoulos1989} which can be appreciated in Fig. \ref{fig:compare_corotation}. The width of the banana orbits, which is related to the maximum value attained by the radial perturbation $r_1$, is dependent on the amplitude of the scalar field but also on the initial angle. The extent of the banana orbit, i.e., the angular range it covers, depends only on the initial conditions of the problem, particularly on the initial angle. Hence, no matter how thin it is, a banana orbit will always be found close to a stable Lagrangian point as a result of an initial angle $\phi_{1i} \neq 0$. On the other hand, the time it takes for an orbiting body to describe a complete banana orbit does not depend on the initial angle, being determined by the mass coupling parameter as 
\begin{equation}
\label{eq:banana_time}
\tau_{\rm banana} \sim \frac{15 M}{(M\mu)^{3}},
\end{equation}
meaning that banana orbits take less time to appear for higher mass couplings\footnote{This timescale is much smaller than the instability time scale of the scalar field we are considering, which is given by $\tau \sim M/(M\mu)^9$~\cite{Brito:2014wla}.}.

If the initial angle is precisely at $\phi_{1i} = \pi/2$ then the resulting orbit is not a banana orbit. This particular initial angle corresponds to an unstable Lagrangian point (see Appendix \ref{app:corotation}) and a particle that starts there will cover the whole angular range. For initial angles bigger than $\pi/2$, the banana orbits are recovered, but in this case they will be centered around the stable Lagrangian point $\phi_i = \pi$.

As this discussion illustrates, one sees that the position of the Lagrangian points determines the shape of corotation orbits. Most notably, the proximity to a given Lagrangian point determines the way the orbiting body reacts to a perturbation: a particle at an unstable Lagrangian point will spend some time at that point, which depends on the mass coupling parameter as
\begin{equation}
  \tau_{\text{unstable}} \sim \tau_{\rm banana} \sim \frac{15 M}{(M\mu)^3},
\end{equation}
and then it will rapidly move to the other unstable Lagrangian point. On the other hand, a particle at a stable Lagrangian point stays there indefinitely or, in case of a perturbation, librates around it - it is this libration that gives rise to banana orbits.

\begin{figure}%
\includegraphics[width=0.85\columnwidth]{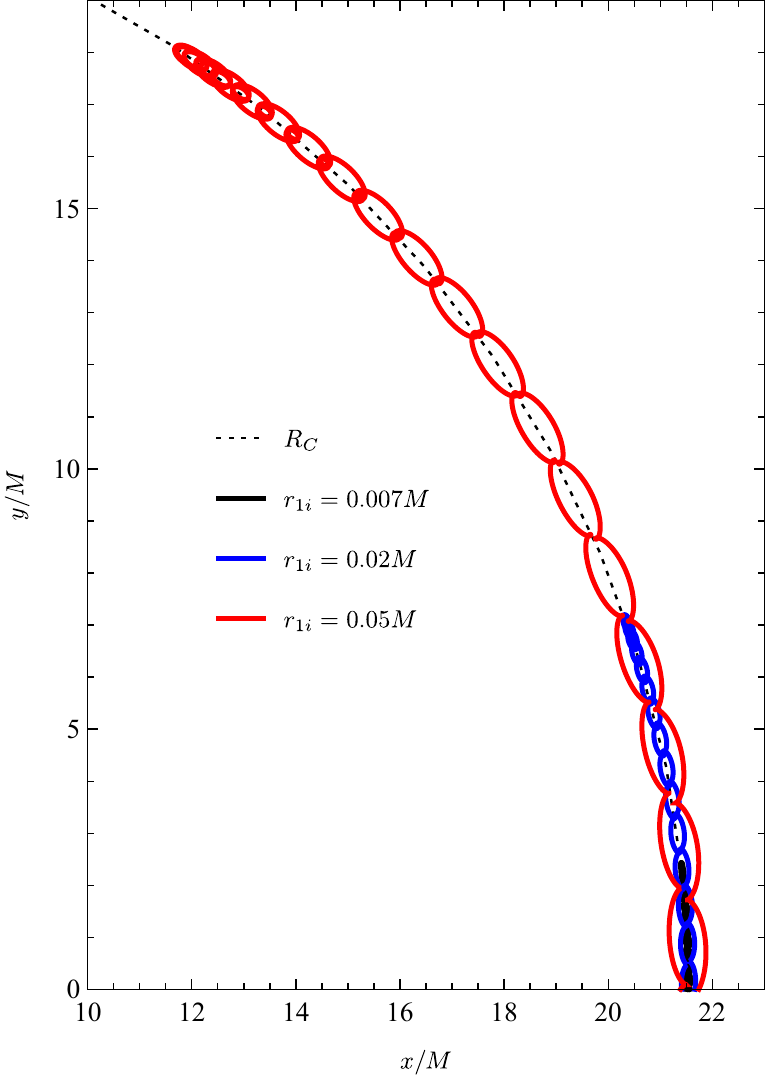}%
\caption{
Orbital motion due to a radial perturbation to a particle at the stable Lagrangian point $\phi_i = 0$. The amplitude of the scalar field is artificially enhanced (we use $\tilde{A}_0 = 350 A_0$) so that the features of the movement are clearer and the mass coupling parameter used is $M\mu = 0.01$. The orbits are initiated at radial position $r(0) = R_C + r_{1i}$ for different values of the radial perturbation $r_{1i}$ as an illustration of the fact that nearly circular orbits in the vicinity of the corotation orbit are described by a libration around stable Lagrangian points. The approximate analytical solution describing the motion is presented in Table~\ref{tab:initial-conditions}.
}
\label{fig:trapping}%
\end{figure}

\begin{table*}[ht]
\centering
\caption{The general solution, described by Eqs.~\eqref{eq:r1corot}-Eq.~\eqref{eq:phi1corot1}, for the specific cases described in Figs.~\ref{fig:compare_corotation} and \ref{fig:trapping}. To obtain these values we applied the method described in Appendix \ref{app:corotation}. From these expressions we can see the immediate influence of the initial perturbations $\phi_{1i}$ and $r_{1i}$ on the amplitude of the solutions.}
\label{tab:initial-conditions}
\begin{tabular}{lll}
\hline
\multicolumn{1}{c}{\textbf{Initial conditions}}         & \multicolumn{1}{c}{$\boldsymbol{r_1(t)}$}                                   & \multicolumn{1}{c}{$\boldsymbol{\phi_1(t)}$}                                \\ \hline
$r_{1i}=\dot{r}_{1i}=\dot{\phi}_{1i} = 0$      & $r_1 = \phi_{1i} [-0.004\sin(0.01 t) + 0.2 \sin(0.0002 t)]$ & $\phi_1 = \phi_{1i} [-0.0004\cos(0.01 t) + \cos(0.0002 t)]$ \\ \hline
$\phi_{1i} = \dot{r}_{1i}=\dot{\phi}_{1i} = 0$ & $r_1 = r_{1i} [-3 \cos(0.01 t) + 4 \cos(0.0002 t)]$         & $\phi_1 = r_{1i} [ 0.3 \sin(0.01 t) + 17 \sin(0.0002 t)]$   \\ \hline
\end{tabular}
\end{table*}
The libration around stable Lagrangian points may induce an accumulation of orbiting bodies in the surrounding regions. In fact, perturbations to a body sitting exactly at a stable Lagrangian point will force it to librate around it, as shown in Fig.~\ref{fig:trapping}, and find itself trapped. A similar effect was obtained in N-body calculations in a galactic setting \cite{ceverino2007} and constitutes a fingerprint of a gravitational potential of the form given in Eq.~\eqref{eq:scalar_grav_potential} -- structures resembling this trapping mechanism close to BHs may be a smoking gun for this dark matter model based on an ultralight scalar field of this sort. 
%
%
\subsection{Orbital torque}

The fact that the perturbing gravitational potential imposed by the presence of the scalar field has an angular component means that the angular momentum is not exactly conserved (cf. Eq.~\eqref{eq:motionphi}). The angular momentum of the orbiting body initialized in a circular orbit of radius $R_0$ is given by
\begin{equation}
  L = (R_0 + r_1)^2 (\Omega_0 + \dot{\phi}_1),
\end{equation}
and the the torque responsible for this is, up to first order, given by
\begin{equation}
  \frac{d L}{dt} = \frac{dL_1}{dt} = 2R_0\Omega_0 \dot{r}_1 + R_0^2\ddot{\phi}_1.
\end{equation}
To get an idea of the magnitude of this effect, we calculate the average value of this quantity over a revolution around the central BH
\begin{equation}
  \left<\frac{dL_1}{dt}\right> = \frac{1}{\Delta t} \int_{0}^{\Delta t} \frac{dL_1}{dt} dt,
\end{equation}
where $\Delta t$ is the interval over which we average. We obtain an expression for this quantity by observing that the expression for $\ddot{\phi}_1$ is related with $r_1$ (see Eq.~\eqref{eq:motionphi2} and \eqref{eq:eom_phi1_corot}) such that we can write
\begin{equation}
  \frac{d L_1}{dt} = 2 P_3 \sin(2(\Omega_0 - \omega_R) t)
\end{equation}
for the general case,
\begin{equation}
  \frac{d L_1}{dt} = 4 P_3 \Big(\sin(2\phi_i) + \cos(2\phi_i) \phi_1 \Big) 
\end{equation}
for the corotation case and
\begin{equation}
  \frac{d L_1}{dt} = \pm 2 P_3 \sin(\kappa_{L\pm} t)
\end{equation}
for the inner and outer Lindblad case. The average values are
\begin{equation}
  \left<\frac{dL_1}{dt}\right> = 0
\end{equation}
for the general case,
\begin{equation}
  \left<\frac{dL_1}{dt}\right> = \pm \frac{(\Omega_0 - \omega_R)}{\pi} P_3 \Bigg( \frac{1 - \cos\Big(\frac{2\pi \kappa_{L\pm}}{\Omega_0-\omega_R}\Big)}{\kappa_{L\pm}} \Bigg)
\end{equation}
for the inner (minus sign) and outer (plus sign) Lindblad resonances, where we used $\Delta t = 2\pi/(\Omega_0-\omega_R)$, and 
\begin{align}
  \left<\frac{dL_1}{dt}\right> &= \frac{2 \omega_R}{\pi} P_3 \frac{}{}\Big( 2\sin(2\phi_i) + 2\cos(2\phi_i) \times \nn\\
  &\Big[\frac{C_3}{\omega_1} \big(1 - \cos(2\pi\omega_1/\omega_R)\big) + \frac{C_4}{\omega_2} \sin(2\pi\omega_2/\omega_R)\Big]\Big)
\end{align}
for the corotation case, where we considered $\Delta t = 2\pi/\omega_R$.

The change in the angular momentum of a particle over a complete orbit around the BH is considerable only when the particle is orbiting at a resonant orbit. Particularly, since the function $P_3$ is overall negative, we see that at the inner Lindblad resonance there is an increase in the angular momentum while at the outer Lindblad resonance there is a decrease. At the corotation resonance the angular momentum transfer depends structurally on the initial angle $\phi_i$: explicitly, as argument of sinusoidal functions, and implicitly, affecting the values of the constants $C_3$ and $C_4$ (cf. Appendix \ref{app:corotation}). The latter fact can be seen by calculating the change in the angular momentum of a particle in a stable Lagrangian point $\phi_i = 0$ or $\phi_i = \pi$; if one would care to go though the calculations one would find that $C_3 = C_4 = 0$ .

In any case, in the mass coupling limit we are considering the angular momentum changes only slightly. Be that as it may, the transfer of angular momentum from the scalar cloud can play an important role on the dynamical evolution of the EMRI in more extreme regimes. In the following we will cover some instances in which similar mechanisms are essential for the dynamics of the systems in which they are inserted.

\subsubsection{Comparison with known phenomena and floating orbits}
In planetary dynamics, particularly in interactions between planetary rings and satellites, angular momentum can be transferred between the disk and the orbiting object~\cite{Goldreich:1980wa,1979ApJ...233..857G,1972MNRAS.157....1L}. This exchange of angular momentum is most effective at the resonances, both corotation and (inner and outer) Lindblad. Angular momentum is removed from the disk when the satellite is at the innermost Lindblad resonance and added to the disk when the satellite is at the outermost Lindblad resonance. Angular momentum is also exchanged at the corotation resonance, in which the rate depends on the gradient of the vorticity per unity of surface density. 
In general, a disk with background (surface) density $\sigma$ has a angular momentum flux induced by the satellite of (dipolar case)
\begin{align}
&\frac{dL}{dt}=\nonumber\\
&-\pi^2\left[\sigma\left(r\frac{dD}{dr}\right)^{-1}\left(r\frac{d\psi_{s}}{dr}+\frac{2 \Omega_d}{\Omega_d -\Omega_m}\psi_s\right)^2\right]_{r_{\rm L}}\,,
\label{eq:angular_transfer}
\end{align}
where the quantity in the brackets are computed at the inner or outer Lindblad resonances, $D=\kappa_d^2-(\Omega_d-\Omega_m)^2$, $\psi_s$ is the gravitational potential generated by the satellite and $\Omega_m$ depends essentially on satellite quantities (see Ref.~\cite{Goldreich:1980wa} for details). At the corotation resonances, the angular momentum flux is given by
\begin{equation}
\frac{dL}{dt}=
\frac{\pi^2}{2}\left[\sigma\left(\frac{d\Omega_d}{dr}\right)^{-1}\frac{d}{dr}\left(\frac{\sigma}{B}\right)\psi_s^2\right]_{r_{\rm C}}\,,
\label{eq:angular_transfer_co}
\end{equation}
where $B=\Omega_d+r\Omega_d'/2$ is the Oort's parameter. Expressions \eqref{eq:angular_transfer} and \eqref{eq:angular_transfer_co} are derived assuming a fluid description for the disk as well as some physical reasonable assumptions for the propagation of the fluid perturbations in the disk.

Although we cannot directly map the system above into the BHSFS, many of its features must be common to the latter. In order to translate the above picture for the BHSFS one has to solve the perturbation equations of the Einstein-Klein-Gordon system around a hairy BH induced by an orbiting particle. 
In general, similarly to the disk-satellite interaction~\cite{1979ApJ...233..857G}, 
there will be modes that are trapped (due to the mass of the scalar field, for instance) and modes that propagate angular momentum and energy away from the system. Since in many scenarios scalar field configurations can be mapped into effective fluid descriptions 
(see Ref.~\cite{Boonserm:2015aqa} for instance), one may expect that equations similar to 
Eqs.~\eqref{eq:angular_transfer} and \eqref{eq:angular_transfer_co} can be obtained from the 
equations governing the perturbations of stars orbiting hairy BHs. Additional mechanisms, 
such as superradiant instability, may also be important, depending on the frequency of the orbiting stellar object. The description of perturbations of hairy BHs with scalar field is still an open subject; to improve it one must go beyond weak field approximations---the perturbations experience strong field regions and appropriate boundary conditions have to be set at the event horizon. Therefore, and though the full description is beyond the scope of this paper, we conjecture that orbital resonances may play an important role in the angular momentum transfer between the scalar field and the orbiting objects in such systems.

The exchange of angular momentum between the stellar object and the scalar field halo can also lead to floating orbits, depending on the configuration of the system. Let us, firstly, note that the EMRI binary formed by a BHSFS and a stellar object incidentally loses angular momentum due to the gravitational radiation emitted. For non-relativistic orbits, the energy and angular momentum flux can be approximated using the quadrupole formalism~\cite{Maggiore:1900zz}. Therefore, the angular momentum flux due to a stellar object in a circular orbit at radius $R$ is given by~\footnote{Gravitational wave emission tends to circularize orbits, and therefore it is sufficient for us to consider only circular orbits here.	}
\begin{equation}
\frac{dL}{dt}=-\frac{32}{5}\frac{\eta^2(M_{\rm total})^{5/2}}{R^{7/2}},
\label{eq:qp_flux}
\end{equation}
where $M_{\rm total}$ is the total mass of the configuration and $\eta$ is the reduced mass. Because the system is losing angular momentum through gravitational waves, the orbit tends to decrease in radius. However, according to the reasoning exposed above, when the system is at the resonances additional angular momentum can be provided to the star [see Eqs.~\eqref{eq:angular_transfer} and \eqref{eq:angular_transfer_co}]. If the angular momentum provided is sufficient to compensate the lost by gravitational waves, the net result is zero and the orbiting object remain with the same angular momentum, even though the system as a whole radiates. This characterizes essentially a floating orbit~\cite{Misner:1972kx,Press:1972zz,Cardoso:2011xi,Fujita:2016yav}. Note that differently to what happens in the standard picture---where the energy is provided by the BH's rotational energy---the energy in the BHSFS is provided by the scalar field. It would be interesting to see whether the order of magnitude between \eqref{eq:angular_transfer} and \eqref{eq:qp_flux} are the same, which favors the above picture. Nevertheless, to achieve such picture, the BHSFS should have the exact configuration such that~\eqref{eq:angular_transfer} equals~\eqref{eq:qp_flux}, which in turns depends on the perturbations of the BHSFS.
\section{Non-minimal effects of scalar fields on orbital motion}\label{sec:forces}

The stellar object orbiting the BHSFS may be subject to other forces besides the standard gravitational one. Here we comment on the possibility of having an additional scalar force, assuming that the stellar matter has a non-minimal coupling with the scalar field. Moreover, we comment on the effects of the dynamical friction and accretion, using a fluid approximation to describe it.

\subsection{Scalar force}
Assuming that the stellar matter is non-minimally coupled with a scalar field, we have that there will be an extra force, which shall depend on the coupling between the scalar field and the stellar matter. A simple way to test the implications of such force, is to assume that the interaction action $S_I$ is such that the motion of the star is described by~\cite{Poisson:2011nh,Burko:2002ge,Warburton:2010eq}
\be
S=S_{0}+S_I=-m_{\rm p}\int d\tau+ q \int d\tau \Phi,
\ee
where $\Phi$ is the (background) scalar field and $q$ measures the strength of the interaction. When $q=0$, the particle follows a geodesic motion. Such kind of coupling may arise in different contexts, such as dimensional reduction or a frame transformation in the action. Using the above action, we find the following equation of motion
\be
\label{eq:gr_scalar_field}
m_{\rm p}(\tau)\frac{D u^a}{d\tau}=q(g^{a b}+u^{a}u^{b})\p_{b}\Phi,
\ee
where $m_{\rm p}(\tau)\equiv m_0-q\Phi$ is usually called dynamical mass~\cite{Poisson:2011nh}, and $D/d\tau\equiv u^a\nabla_a$. The above equation of motion can also be obtained from the action~\cite{Quinn:2000wa}
\be
S=\int d^4x \delta^4(x-z(\tau)) \l[\frac{1}{2}m_{\rm p}\,g_{ab}u^{a}u^b+q\Phi\r]\,,
\ee
where $z(\tau)$ represents the worldline of the particle. We shall use the above setup to determine the features of a force due to a background scalar field. 
This setup may suit the description of stellar-size objects through clouds surrounding BHs~\cite{Benone:2014ssa,Herdeiro:2014goa,Herdeiro:2015waa}, through dark matter mini-spikes~\cite{Eda:2013gg,Eda:2014kra} or even through compact exotic objects formed by scalar field, such as boson stars~\cite{Schunck:2003kk,Macedo:2013jja,Macedo:2013qea}. Note that the above picture is fully relativistic.

\begin{figure}%
\includegraphics[width=0.8\columnwidth]{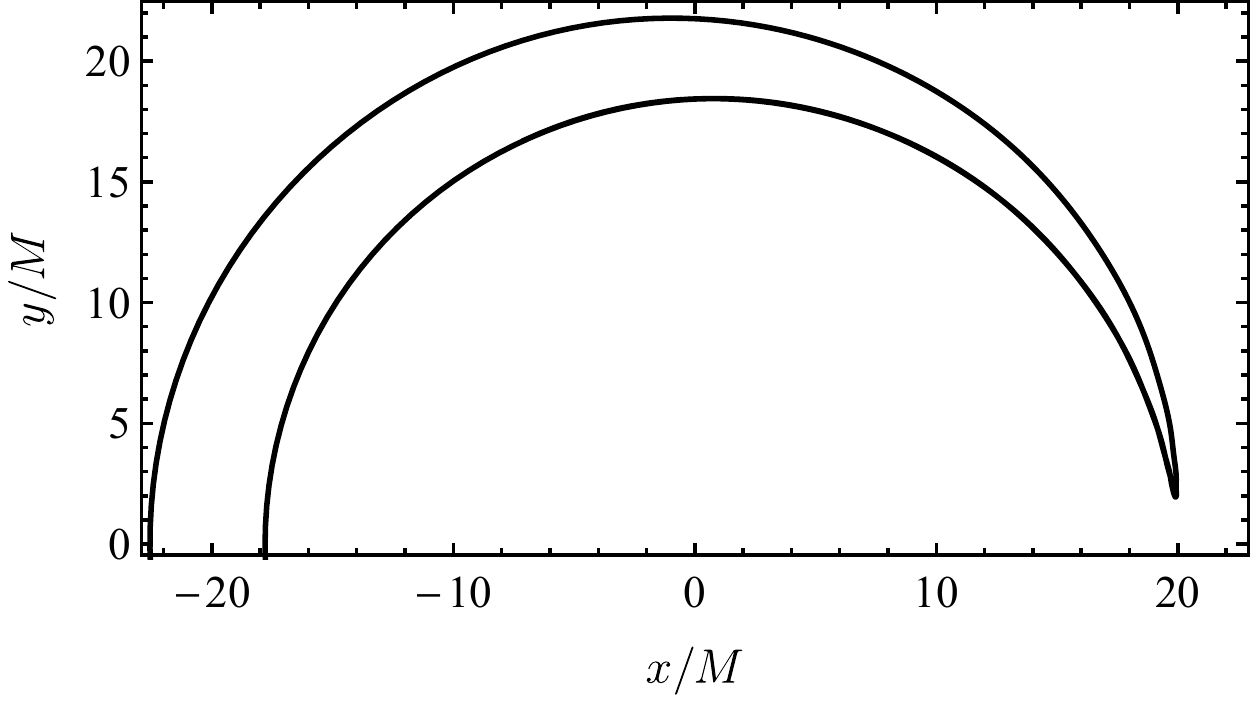}%
\caption{Illustrative example of banana orbits considering only the scalar force, obtained by solving the equation for a charged scalar particle~\eqref{eq:gr_scalar_field}, in the Schwarzschild background and the scalar field profile used in this paper. The plot is symmetric with respect to the $x$-axis.}%
\label{fig:scalar_ban}%
\end{figure}

Let us begin, for simplicity, neglecting the gravitational interactions between the scalar field and the star. Therefore, the star will feel the BH spacetime and the additional scalar force will be weighted by the coupling parameter $q/m_{\rm p}$. Assuming a dependence of the field as $\Phi=\psi(r)\cos (\phi-\omega_R t)$, and considering a small $q$ limit, all quantities can be expanded around the geodesic quantities, just as seen in previously in Sec.~\ref{sec:quasicircular}. In this case, considering that at zeroth order the particle is at a circular orbit and that the corrections are small, we have that the motion can be described as
\bea
t&=&\dot{t}_0 \tau+t_1(\tau),\nonumber\\
r&=&R+r_1(\tau),\label{eq:approx}\\
\phi&=&\dot{\phi}_0 \tau+\phi_1(\tau),\nonumber
\eea
where $\dot{t}_0$ and $\dot{\phi}_0$ are constants obtained from solving the zeroth order problem, and we have chosen that the motion initially starts at $t(0)=0$ and $\phi(0)=0$. Using this solution, the equations for $t_1$, $\phi_1$, and $r_1$ and are given by
\begin{widetext}
\begin{align}
&\ddot{t}_1-\frac{q \psi _0 \left(\dot{t}_0 \dot{\phi }_0 (R_0-2 M)+\dot{t}_0^2 \omega_R  (2 M-R_0)+R_0 \omega_R \right) \sin \left[( \dot{t}_0 \omega_R - \dot{\phi }_0)\tau\right]}{m_0 (R_0-2 M)}+\frac{2 M \dot{r}_1 \dot{t}_0}{R_0{}^2-2 M R_0}=0,\label{eq:t1eq}\\
&\ddot{\phi }_1-\frac{q \psi _0 \left(R_0{}^2 \dot{\phi }_0^2+1-R_0{}^2 \dot{t}_0 \omega_R  \dot{\phi }_0\right) \sin \left[( \dot{t}_0 \omega_R - \dot{\phi }_0)\tau\right]}{m_0 R_0{}^2}+\frac{2 \dot{r}_1 \dot{\phi }_0}{R_0}=0,\label{eq:phi1eq}\\
&\ddot{r}_1+\frac{q (R_0-2 M) \left(\psi _0 \left(R_0{}^3 \dot{\phi }_0^2-M \dot{t}_0^2\right)-R_0{}^2 \partial _r\psi _0\right) \cos \left[( \dot{t}_0 \omega_R - \dot{\phi }_0)\tau\right]}{m_0 R_0{}^3}+\frac{2 M \dot{t}_1 \dot{t}_0 (R_0-2 M)}{R_0{}^3}\nonumber\\
&+\dot{\phi }_1 \left(4 M \dot{\phi }_0-2 R_0 \dot{\phi }_0\right)+r_1 \left(-\frac{2 M \dot{t}_0^2 (R_0-3 M)}{R_0{}^4}-\dot{\phi }_0^2\right)=0,
\end{align}
\end{widetext}
where we have defined $\psi_0\equiv \psi(R_0)$. The first two equations can be directly integrated and solved for $\dot{t}_1$ and $\dot{\phi}_1$. We then use the result into the radial equation, obtaining
\be
\ddot{r}_1+\kappa^2 r_1={\cal F}(R_0)\cos[( \dot{\phi}_0-\omega_R \dot{t}_0)\tau],
\label{eq:ddr_scalar}
\ee
where $\kappa$ is the usual epicyclical frequency, measured according to the particle's proper time, and 
\begin{align}
{\cal F}\equiv&- \frac{2 q \psi _0 ( R_0 \Omega_0-M (2 \Omega_0+\omega_R ))}{m_0 R_0^2 (\omega_R - \Omega_0)}\nonumber\\
&-\frac{q \partial _r\psi _0 (R_0-2 M)}{m_0 R_0}.
\end{align}
Note that the above equation is very similar to Eq.~\eqref{eq:eq_ddr}. This suggests that the same effects appearing in the gravitational field due to a scalar field also appear when the cloud itself interacts directly with the star. However, note that the angular dependence of the gravitational external force is twice the one of the scalar external force. This enables longer banana orbits, which can be seen in the interval $0<\phi<2\pi$, as illustrated in Fig.~\ref{fig:scalar_ban}. The width of the banana orbits also depends on the coupling $q/m_{\rm p}$. Additionally, note that Eq.~~\eqref{eq:ddr_scalar} is relativistic, unlike the gravitational case we explored previously, and therefore it is valid for \textit{any} orbit; once again, one must be careful at resonant configurations, mainly because the approximations given by Eq.~\eqref{eq:approx} breaks down.
%

\subsection{Gravitational drag and accretion}

We can get a grasp into additional effects of the scalar field by analyzing some known features of gravitating objects moving through fluids. Here we briefly describe two of them: accretion and dynamical friction. The expressions are valid in the Newtonian regime.

Accretion causes the stellar object to increases its mass, therefore changing its momentum. The accretion rate is usually described by~
\footnote{Note that we are using an expression for wind accretion, which is reasonable because the characteristic size of the stellar object is much smaller than the environment.}
\be
\dot{m}_0=\sigma \rho v\,,
\ee
where $\sigma$ is the accretion cross section, that depends on the nature of process, $\rho$ the local density and $v$ the relative velocity between the fluid and the object. The simplest accretion mode, suitable for our scenario, describes the accretion of colisionless particles. For this case, the accretion rate is~\cite{Unruh:1976fm,shapiro,Giddings:2008gr}
\be
\dot{m}_0=\frac{\pi \rho R_p}{v},
\ee
where $R_p$ is the characteristic size of the stellar object. The accretion into the moving star effectively creates a force in the opposite direction to the velocity of the star.

Besides accretion, the interaction between the stellar object and the wake left behind can be important. Assuming that sound speed in the medium to be large compared to the one of the stellar object, we can model the gravitational friction as~\cite{Ostriker:1998fa,Kim:2007zb,Macedo:2013qea}
\be
{\bf F}_{\rm DF}=-\frac{4\pi m_0\rho}{v^2}\l[\frac{1}{2}\ln\l(\frac{1+v/c_s}{1-v/c_s}\r)-v/c_s\r]{\bf \hat v}.
\ee
For the cases explored here, the velocity of the scalar medium is very small. Also, the propagation of scalar perturbations is in general very large compared with the relative velocity $v$. Therefore, a good approximation for the drag force is given by
\be
{\bf F}_{\rm DF}\approx -\frac{4\pi }{3}m_0\rho_1{\bf v},
\label{eq:df_force}
\ee	
where we have used that $c_s\sim 1$.

\begin{figure}%
\includegraphics[width=0.8\columnwidth]{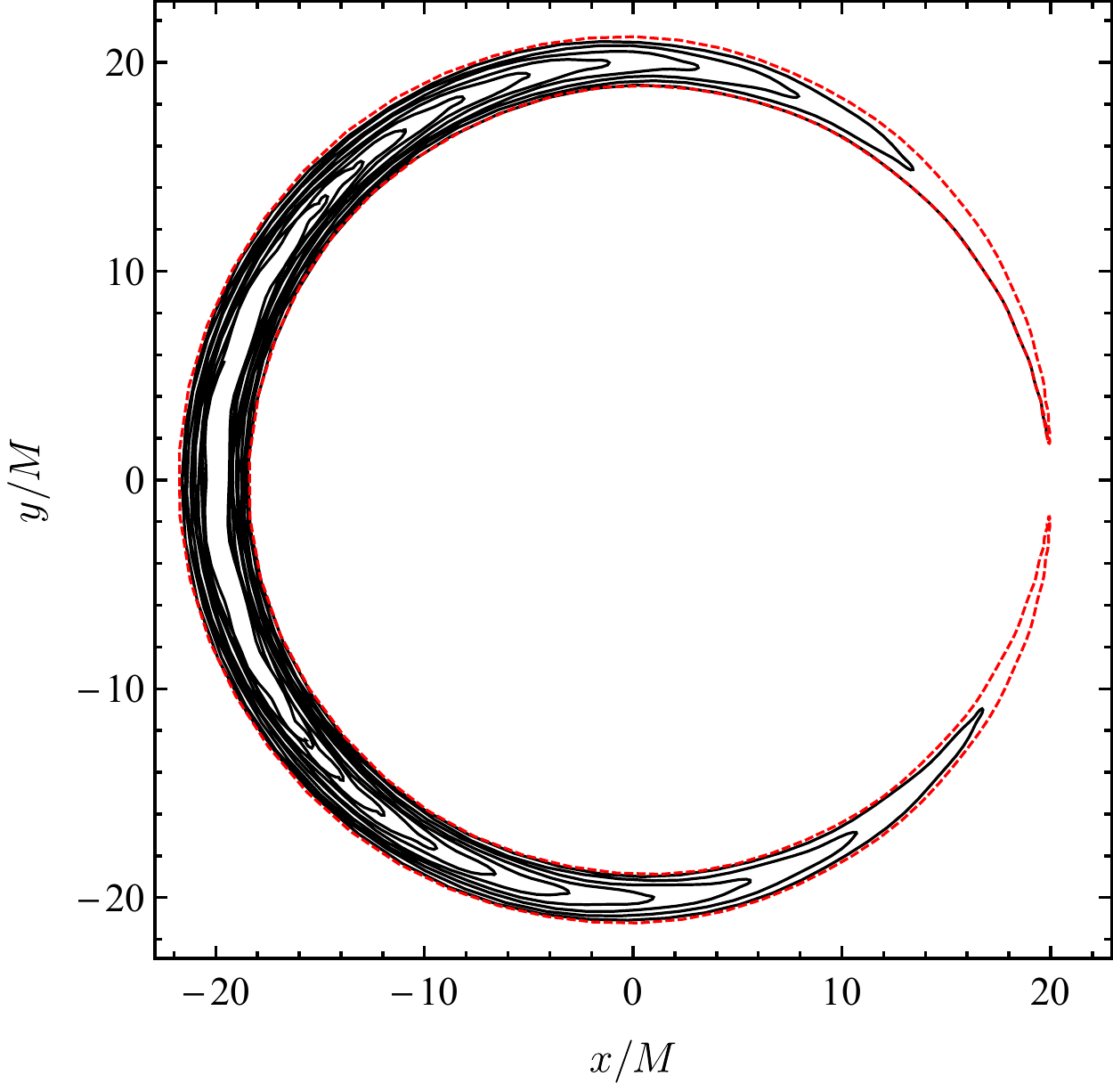}%
\caption{Near corotating orbit, considering a simple dynamical friction term proportional to the velocity. The dotted line is the result considering no drag force. We see that the effect of dragging forces is the shrink the banana orbit, making the stellar object closer to the Lagragian point.}%
\label{fig:scalar_ban2}%
\end{figure}

We see that both accretion and dynamical friction create a dragging force, in the opposite direction to the velocity of the moving object. In order to see the influence of the dragging forces into the motion of the scalar charge, we artificially insert the expression of an external force $-b{\bf v}$ into the relativistic equation of the motion, with $b$ being a dimensionful constant, suitably chosen to have a very small scaled value. While this is not valid rigorously, it may give a hint about the effects of accretion and dynamical friction. The result can be seen in Fig.~\ref{fig:scalar_ban2}. The effect of the dragging force is to shrink the size of the banana orbit, which makes the particle approach the stable Lagrangian point. We note that in the case we have only a scalar force in a BHSFS, therefore neglecting the gravitational force of the scalar field, there is only one stable Lagragian point located at the corotation radius and $\phi=\pi$.

%

\section{Discussion and final remarks} \label{sec:discusscf}

The existence of new, ultralight fundamental fields are an intriguing prospect: in the many orders of magnitude ranging between the lightest known particle and the cosmological horizon scale, new physics can hide. One natural possibility is to populated this desert with light fields. These fields could, in addition, explain dark matter, the accelerated expansion of our universe, or just work as a toy model for more complex interactions. 

A generic outcome of having massive bosons living together with supermassive BHs, is the spin-down of the BH and the growth of a bosonic, non-axisymmetric structure~\cite{Brito:2017wnc,Brito:2015oca}. This bosonic structure may give rise to measurable gravitational waves, or lead to such a low BH-spin that it would be noticeable in a statistic sense~\cite{Brito:2017wnc,Brito:2017zvb}.
Our results are sensitive to the mass parameter $\mu$, which controls the scales at which the the scalar field effects play a role; systems with a different value of the coupling $M\mu$ will show a different phenomenology at different scales. Indeed, a scalar field with mass $\mu$ will produce effects whose scales will ultimately depend on the central mass $M$. Fortunately, BHs come in a wide mass-range: from stellar-mass BHs to $10^6M_{\odot}$ in the center of Milky Way \cite{0004-637X-689-2-1044}, and to $10^{10}M_{\odot}$ solar masses in the center of the galaxy NGC 4889 \cite{mcconnell2011two}. Thus, one has, in principle, scalars with a mass scale differing by several orders of magnitude may give rise to similar effects, albeit around BHs of different masses.

What we have shown here is that there is a direct imprint of bosonic structures in the orbit of tightly-bound stars. The asymmetric bosonic structure leads to periodic forcing of nearby stars,
which triggers resonances or anomalous precession effects. Overall, the bosonic structure may cause the BH -- star-populated-vicinity to behave like large-scale galactic structure. A mapping of the star content close to supermassive BHs may reveal the presence of such structures; in the meanwhile, an extension of all our results to the relativistic regime is clearly necessary.

\begin{acknowledgements}
We would like to thank Enrico Barausse, Sante Carloni, Alexandre Correia, Viktor Czinner, Carlos Herdeiro and Paolo Pani, for fruitful discussions.

The authors acknowledge financial support provided under the European Union's H2020 ERC Consolidator Grant ``Matter and strong-field gravity: New frontiers in Einstein's theory'' grant agreement no. MaGRaTh--646597. 
Research at Perimeter Institute is supported by the Government of Canada through Industry Canada and by the Province of Ontario through the Ministry of Economic Development $\&$
Innovation.
C.M. thanks the support from Conselho Nacional de Desenvolvimento Cient\'ifico e Tecnol\'ogico (CNPq) and Coordenação de Aperfeiçoamento de Pessoal de Nível Superior (CAPES).
M. F. acknowledges financial support provided by Funda\c{c}\~{a}o para a Ci\^{e}ncia e a Tecnologia Grant number PD/BD/113481/2015 awarded in the framework of the Doctoral Programme IDPASC - Portugal.
This project has received funding from the European Union's Horizon 2020 research and innovation programme under the Marie Sklodowska-Curie grant agreement No 690904.
This article is based upon work from COST Action CA16104 ``GWverse'', supported by COST (European Cooperation in Science and Technology).
The authors thankfully acknowledge the computer resources, technical expertise and assistance provided by S\'ergio Almeida at CENTRA/IST. Computations were performed at the cluster ``Baltasar-Sete-S\'ois'', and supported by the MaGRaTh--646597 ERC Consolidator Grant.
\end{acknowledgements}

\appendix

\section{Weak-field analysis}\label{app:weak-field}

In the limit of small mass-coupling parameter, $M\mu \ll 1$, the peak of the scalar field profile is localized far from the BH, which enables the use of a flat background in its analysis (see \cite{Brito:2015oca,Yoshino01042014}). The gravitational effect of the presence of the scalar field is obtained by studying the linear response of the metric to its stress-energy tensor. We shall follow the approach presented in \cite{PhysRevD.87.044017}. We consider that the spacetime metric is written as
\begin{equation}
g_{\mu\nu} = \eta_{\mu\nu} + h_{\mu\nu},
\end{equation}
and that, to linear order, the interaction between the scalar field and the metric is governed by the action
\begin{equation}
\label{eq:linear-action}
S = \int d^4x \left[ \frac{1}{2\kappa^2} h_{\mu\nu} \mathcal{E}^{\mu\nu,\rho\sigma} h_{\rho\sigma} + \frac{1}{2} h_{\mu\nu} T^{\mu\nu} \right],
\end{equation}
where $\kappa \equiv (32 \pi)^{1/2}$, $\mathcal{E}^{\mu\nu,\rho\sigma}$ is the Lichnerowicz operator and 
\begin{equation}
T_{\mu\nu} = \frac{1}{2} \left[ \Phi^*_{,\mu} \Phi_{,\nu} + \Phi^*_{,\nu} \Phi_{,\mu} - \eta_{\mu\nu} (\eta^{\rho \sigma} \Phi^*_{,\rho} \Phi_{,\sigma} + \mu^2 |\Phi|^2) \right]
\end{equation}
is the stress-energy tensor of the scalar field.
Notice that the to raise and lower indices we use the background Minkowski metric. To obtain the equations of motion for the metric components, it is useful to perform the scalar-vector-tensor decomposition\footnote{A symmetric $4\times 4$ spacetime tensor $A_{\mu\nu}$ in a constant curvature background manifold can be decomposed into scalar, vector and tensor independent parts. This is the statement of the scalar-vector-tensor decomposition. This decomposition is related to the way each component transforms under the group of rotations of the background. Given that $A_{00}$ has no spatial indices, it is a scalar; $A_{0i} = A_{i0}$ has one spatial index, so it is a vector; finally, $A_{ij} = A_{ji}$ is a spatial tensor. The decomposition is not complete since it is possible to decompose both the vector $A_{0i}$ and the tensor $A_{ij}$ in scalar, vector and tensor components. This second part of the decomposition is made by observing that the symmetry of the spatial part of the background manifold allows the expansion of vectors and symmetric tensors in terms of solutions of the Helmholtz equation (see, for instance,  \cite{kodama1984cosmological,Bertschinger:1993xt,bardeenCOSMO,Maggiore:1900zz,Czinner2012}). This is achieved by projecting each object in the basis elements of the space of solutions of the Helmholtz equation \cite{Czinner2012,Pettinari:2014vja}. Consequently, an arbitrary vector is broken into a scalar and a transverse vector (a rendition of the original Helmholtz theorem); an arbitrary symmetric tensor is decomposed into two scalars, one transverse vector and a transverse traceless tensor \cite{Pettinari:2014vja,Maggiore:1900zz}.} 
of both the perturbation tensor $h_{\mu\nu}$
\begin{align}
h_{00}= & 2 \psi\,,\label{eq:metric-decomp1}\\
h_{0i} = & \beta_i + \partial_i \gamma\,,\label{eq:metric-decomp2}\\
h_{ij} = & - 2\phi \delta_{ij} + (\partial_i \partial_j - \frac{1}{3} \delta_{ij} \nabla^2)\lambda + \nn\\
&\frac{1}{2} (\partial_{i}\epsilon_{j} + \partial_j \epsilon_i) + h_{ij}^{TT}\,\label{eq:metric-decomp3},
\end{align}
and the stress-energy tensor
\begin{align}
T_{00} =& \rho\,,\label{eq:stressDECOMP1}\\
T_{0i} =& S_i + \partial_i S\,,\label{eq:stressDECOMP2}\\
T_{ij} =& P \delta_{ij} + (\partial_i \partial_j - \frac{1}{3} \delta_{ij} \nabla^2) \sigma +\nn\\
& \frac{1}{2}(\partial_i \sigma_j + \partial_j \sigma_i) + \sigma_{ij}\,,\label{eq:stressDECOMP3}
\end{align}
where $\nabla^2$ is the flat space Laplacian,  $\psi, \phi,\gamma,\lambda$ are scalars under spatial rotations, $\beta^i$ and $\epsilon^i$ are transverse vector fields and $h_{ij}^{TT}$ is a transverse traceless $3\times3$ tensor;  the components of the stress-energy tensor have, \textit{mutatis mutandis}, the same meaning as their metric counterparts. Additionally, the decomposed components satisfy
\begin{equation}
\partial_i \beta^i = 0, \quad \partial_i \epsilon^i = 0, \quad \partial^j h_{ij}^{TT} = 0, \quad \delta^{ij} h_{ij}^{TT} = 0,
\end{equation}
for the metric ones, and similarly for the stress-energy tensor ones. This decomposition is unique as long as boundary conditions guaranteeing asymptotical flatness are satisfied \cite{PhysRevD.87.044017}.

Of the ten degrees of freedom of the metric, only six of them are gauge invariant. These degrees of freedom are constructed from the decomposed components of the metric in Eqs. \eqref{eq:metric-decomp1} - \eqref{eq:metric-decomp3} and constitute a particular case of the more general Bardeen variables \cite{bardeenCOSMO}
\bea
  \tilde{\Phi} = - \phi - \frac{1}{6} \nabla^2 \lambda\,,\label{eq:bardeenVAR1}\\
  \tilde{\Psi} = \psi - \dot{\gamma} + \frac{1}{2} \ddot{\lambda}\,,\label{eq:bardeenVAR2}\\
  \Xi_i = \beta_i - \frac{1}{2} \dot{\epsilon}_i\label{eq:bardeenVAR3}.
\eea
$\tilde{\Phi}$ and $\tilde{\Psi}$ account for two degrees of freedom, $\Xi$, which is subjected to the condition $\partial_i \Xi^i = 0$, carries two degrees of freedom and the remaining two are carried by $h_{ij}^{TT}$ which, by construction, is gauge invariant.
Writing the linearized action \eqref{eq:linear-action} in terms of the gauge-independent variables and the decomposed components of the stress energy tensor, one arrives at the equations of motion
\bea
  \nabla^2 \tilde{\Phi} = - 4\pi\rho\,,\\
  \nabla^2 \tilde{\Psi} = -4 \pi (\rho + 3P - 3 \dot{S})\,,\\
  \nabla^2 \Xi_i = - 16\pi S_i\,,\\
  \square h_{ij}^{TT} = -16 \pi \sigma_{ij}\,,
\eea
where $\square = \eta^{\mu\nu} \partial_{\mu} \partial_{\nu}$. This means that, at linear order, the different components of the metric decouple, which allows one to treat them independently. In what follows, we shall focus on the scalar components of the metric.

\subsection{Gauge Choice}
As it was mentioned previously, of the ten degrees of freedom carried by the metric, four of them are non-physical because they are related with the gauge freedom of the theory. So far, we have followed a gauge independent analysis of the equations in order to focus solely on the physically relevant, gauge independent, quantities. This is not the only way of avoiding the gauge dependent quantities in the calculations: one can also fix a gauge, i.e., one can impose conditions on the variables of the problem in order to remove the non-physical degrees of freedom. The Newtonian or longitudinal gauge \cite{MUKHANOV1992203} is defined by imposing the conditions
\begin{equation}
  \beta_i = 0, \quad \gamma = 0, \quad \lambda = 0, \quad \epsilon_i = 0, \quad h_{ij}^{TT} = 0,
\end{equation}
which means that not only the four gauge degrees of freedom are being turned off, but also four physical degrees of freedom are being explicitly discarded. Hence, given that we are not interested in these degrees of freedom, this metric is ideal for what follows, since it only conveys the scalar metric perturbations\footnote{Other works have employed this gauge - see \cite{Khmelnitsky:2013lxt} and \cite{Blas:2016ddr}}. In this gauge, the line element is
\begin{equation}
  ds^2 = -(1 - 2\psi) + (1-2\phi)\delta_{ij} dx^i dx^j;
\end{equation}
where it was made the identification between the scalar components $\phi$ and $\psi$ of Eqs.\eqref{eq:metric-decomp1} and \eqref{eq:metric-decomp3} and the gauge invariant scalars $\tilde{\Phi}$ and $\tilde{\Psi}$ of Eqs. \eqref{eq:bardeenVAR1} and \eqref{eq:bardeenVAR2} given that in this gauge they are equal \cite{MUKHANOV1992203,Bertschinger:1993xt}. Moreover, the equations of motion governing the scalar potentials $\phi$ and $\psi$ are equal \cite{MUKHANOV1992203} to the ones governing $\tilde{\Phi}$ and $\tilde{\Psi}$, completing the identification.
\newline

\subsection{Particle motion in a weak gravitational field}
The Lagrangian that describes the movement of a free particle is given by
\begin{equation}
  \mathcal{L} = \sqrt{- g_{\mu\nu} dx^{\mu} dx^{\nu}}
\end{equation}
which is written, in the Newtonian gauge and up to first order in the scalar potentials, as \cite{Ma:1993xs}
\begin{equation}
  \mathcal{L} = \sqrt{1-v^2} \left( 1 - \frac{\psi - \phi v^2}{1-v^2}\right),
\end{equation}
where $v^2 = \dot{x}^i\dot{x}^j \delta_{ij}$ is the coordinate velocity of the particle measured by a particular observer that sits far enough from the central BH. To write the Euler Lagrange equations we need
\begin{equation}
  \frac{\partial \mathcal{L}}{\partial \dot{x}^i} = \frac{\dot{x}^i}{\sqrt{1 - v^2}} \left(-1+2\phi - \frac{\psi - \phi v^2}{1 -v^2} \right),
\end{equation}
and
\begin{equation}
  \frac{\partial \mathcal{L}}{\partial x^i} = \frac{-1}{\sqrt{1 - v^2}} (\partial_i\psi - (\partial_i \phi)v^2).
\end{equation}
To obtain these expressions we considered that we are in a weak field limit, keeping only the first order terms in the potentials. On top of this, we shall consider that the particle is non-relativistic and that the time derivatives of the potentials are very small, i.e. that a quasi-static limit applies. Taking this into account, all the terms proportional to the velocity $v$ can be ignored, as well as the terms $\dot{x}^i \phi$ and $\dot{x}^i \psi$. Finally, the equation of motion for a free, non-relativistic particle in a a quasi-static, weak field limit is given by
\begin{equation}
  \frac{d}{dt}\left( \frac{\partial \mathcal{L}}{\partial \dot{x}^i} \right) = \frac{\partial \mathcal{L}}{\partial x^i} \Rightarrow \ddot{x}^i = \partial_i\psi
\end{equation}
and we have the identification of the scalar potential $\psi$ with the Newtonian potential due to the presence of the scalar field, indicated as $\Psi_1$ in Eq,~\eqref{eq:scalar-field-potential-schem}.

Following the previous considerations, we shall study the influence of the scalar field in the movement of a point particle through the use of a Newtonian gravitational potential $\Psi_1$ obtained from the equation
\begin{equation}
  \nabla^2 \Psi_1 = - 4\pi (\rho + 3P - 3 \dot{S})
 \end{equation}
where $\rho$, $P$ and $S$ are given in Eqs.~\eqref{eq:stressDECOMP1}, \eqref{eq:stressDECOMP2} and \eqref{eq:stressDECOMP3}.

\section{Gravitational field of the scalar cloud}\label{app:potential}
\subsection{Harmonic expansion}
%
We obtain the expression for the potential $\Psi_1(t,r,\phi,\theta)$ by employing a method of decomposition of the components of the function in terms of spherical harmonics $Y_{\ell m}(\theta, \phi)$. We follow chapter 1 of Ref.~\cite{Poisson:1639582}. We consider that, being a solution of the Poisson's equation, $\Psi_1$ has an expansion of the form
\begin{equation}
  \label{eq:series-solution}
  \Psi_1 = \sum_{\ell m} \frac{4\pi}{2\ell +1} \left[q_{\ell m}(t,r)\frac{Y_{\ell m}(\theta,\phi)}{r^{\ell + 1}} + p_{\ell m}(t,r) r^{\ell} Y_{\ell m}(\theta, \phi) \right]  
\end{equation}
with
\begin{align}
  q_{\ell m}(t,r) = \int_0^r s^{\ell} \tilde{\rho}_{\ell m}(t,s) s^2 ds\\
  p_{\ell m}(t,r) = \int_r^{\infty} \frac{\tilde{\rho}_{\ell m}(t,s)}{s^{\ell +1}} s^2 ds
\end{align}
and
\begin{equation}
  \tilde{\rho}_{\ell m}(t,r) = \int \tilde{\rho}(t,r,\theta,\phi) Y^*_{\ell m}(\theta,\phi) d\Omega
\end{equation}
where $\tilde{\rho}$ is the source of the potential; in this case the source of the potential is $\tilde{\rho} = (\rho + 3P - 3 \dot{S})$ (see Eqs.~\eqref{eq:stressDECOMP1}, \eqref{eq:stressDECOMP2} and \eqref{eq:stressDECOMP3}).

All the non-zero terms of the expansion of Eq.~\eqref{eq:series-solution} exist up to $\ell = 2$ such that the potential can be written as
\begin{equation}
\Psi_1 = P_1(r) + P_2(r)\cos^2(\theta) + P_3(r) \sin^2(\theta) \cos(2(\phi -\omega_R t))
\end{equation}
where
\begin{widetext}
\begin{align}
P_1(r) =& \frac{A_0^2 \pi e^{-Mr\mu^2}}{2M^5 r^3\mu^8} \Bigl( -192 -192 M r \mu^2 + 2 M^6 r^4 \mu^{10} + M^7 r^5 \mu^{12} - 4 M^5 r^3 \mu^8 (-3+r^2\mu^2)  - 24M^4 r^2 \mu^6 (-1 + r^2\mu^2)  \nn\\
&  + M^2(16 \mu^2 - 160 r^2 \mu^4) + M^3(16 r \mu^4 - 80 r^3 \mu^6) - 16 e^{Mr\mu^2} (-12 + M^4 r^2 \mu^6 + M^2 (\mu^2 - 4 r^2 \mu^4)) \Bigl),\\
P_2(r) =& \frac{A_0^2 \pi e^{-Mr\mu^2}}{2M^5 r^3\mu^8} \Bigl( 576 + 576 M r \mu^2 - 2M^6 r^4 \mu^{10} - M^7r^5\mu^{12} + 48 e^{Mr\mu^2} (-12 + M^2 \mu^2) + 4 M^5 r^3 \mu^8 (-2 + r^2\mu^2) \nn\\
&  + 24 M^4 r^2\mu^6 (-1 + r^2 \mu^2) + 48 M^2 \mu^2 (-1 + 6r^2\mu^2) - M^3(48 r \mu^4 - 96 r^3 \mu^6) \Bigl), \\
  P_3(r) =& \frac{ A_0^2 \pi e^{-Mr\mu^2}}{2M^5 r^3\mu^8} \Bigl(-3456 - 3456 M r \mu^2 - 2M^6 r^4 \mu^{10} - M^7 r^5 \mu^{12} + 48 e^{Mr\mu^2} (72 + M^2 \mu^2) - 8 M^5 r^3 \mu^8 (1 + 3r^2\mu^2) \nn\\
  &  - 48 M^2 (\mu^2 + 36 r^2 \mu^4) - 48 M^3 (r\mu^4 + 12 r^3 \mu^6) - 24 M^4 (r^2\mu^6 + 6 r^4 \mu^8) \Bigl).
\end{align}
\end{widetext}

In the main text, the focus is on the potential in the equatorial plane, $\theta = \pi/2$, which is written as
\begin{equation}
\Psi_1 = P_1(r) + P_3(r) \cos(2(\phi -\omega_R t)).
\end{equation}

\subsection{The Lagrangian points}
A general potential $\Psi = \Psi(r,\phi)$ produces a motion governed by equations on a plane $(r,\phi)$ rotating with angular velocity $\Omega_p$ given by
\begin{align}
  \ddot{r} - r(\dot{\phi} + \Omega_p)^2 + \frac{\partial \Psi}{\partial r} = 0\\
  \frac{d}{dt} ( r^2 (\dot{\phi} + \Omega_p)) + \frac{\partial \Psi}{\partial \phi} = 0
\end{align}
The Lagrangian points are the points where the forces acting on the orbiting particle cancel exactly. To uncover those locations, one forces the equations of motion to describe a particle at rest in this frame, i.e. $\dot{r} = \ddot{r} = \ddot{\phi} = \dot{\phi} = 0$, which amounts to
\begin{align}
  \frac{\partial \Psi}{\partial r} &= r\Omega_p^2, \label{eq:lagrangian-p-radial}\\
  \frac{\partial \Psi}{\partial \phi} &= 0. \label{eq:lagrangian-p-angular}
\end{align}

Applying this reasoning to the total potential in Eq.~\eqref{eq:total-potential}, $\Psi = \Psi_0 + \Psi_1$, one can see from Eq.~\eqref{eq:lagrangian-p-angular} that the Lagrangian points are located at $\phi = 0, \pi/2,\pi, 3\pi/2,...$ since
\begin{equation}
 \frac{\partial \Psi}{\partial \phi} = 0 \Leftrightarrow \sin(2\phi) = 0 \Leftrightarrow \phi = 0, \pi/2, \pi, 3\pi/2,...
\end{equation}
Substituting these values in Eq.~\eqref{eq:lagrangian-p-radial}, we obtain that the radial position of the Lagrangian points satisfies
\begin{equation}
\frac{\partial \Psi_0}{\partial r} + \frac{\partial P_1}{\partial r} \pm \frac{\partial P_3}{\partial r} =  r \Omega_p^2,
\end{equation}
where $\pm$ refers to the unstable ($\phi = \pi/2,...$) or stable points ($\phi = 0,...$), respectively. Considering that the derivatives of both $P_1$ and $P_3$ are negligible, which is a safe assumption in general (see Sec.~\ref{subsubsec:circular orbits}), we obtain that the radial location of the Lagrangian points is given by
\begin{equation}
  \frac{1}{r} \frac{d \Psi_0}{dr} - \Omega_p^2 = 0 \Leftrightarrow \Omega(r)^2 - \Omega_p^2 = 0
\end{equation}
which means that these points are located in a circle with radius given by the radius of the Keplerian circular orbit with angular velocity equal to $\Omega_p$. Given that this is the velocity at which the reference frame is rotating, this is called the corotation radius.

\section{Analytical expression for the perturbation to the circular orbit at corotation} \label{app:corotation}

In this appendix we present some details regarding the analytical solution for the perturbations to the circular orbit at corotation. The equations of motion \eqref{eq:eom_r1_corot} and \eqref{eq:eom_phi1_corot} can be written as
\be
\frac{d X}{dt}  = \hat{A} X + B,
\ee
in which
\be
X^T = (r_1,\phi_1,R_1,\Phi_1),
\ee
with $R_1 = \dot{r}_1$, $\Phi_1 = \dot{\phi}_1$,
\begin{equation}
\hat{A} =
\begin{pmatrix}
0 & 0 & 1 & 0\\
0 & 0 & 0 & 1\\
-(\Psi_0^{''} -\omega_R^2) & -2 P_3'\sin(2\phi_i) & 0 & 2 \omega_R R_c\\
0 & \frac{4 P_3}{R_C^2} \cos(2\phi_i) & - \frac{2 \omega_R}{R_C}  & 0
\end{pmatrix},
\end{equation}
and
\be
B^T = \bigg(0,0,-C(R_C) - P_3' \cos(2\phi_i) ,\frac{2 P_3}{R_C^2} \sin(2\phi_i) \bigg).
\ee
The solution, obtained from standard methods, has the form
\be
X = X_h + X_p,
\ee
in which
\be
X_h = \sum_{i=1}^{4} c_i V_i \exp(\lambda_i t),
\ee
with $c_i, V_i, \lambda_i$ being constants of integration, eigenvectors and eigenvalues of $\hat{A}$, respectively, and $X_p$ is a constant vector. 

The general form of the solutions will depend on Lagrangian point around which the analysis is being made. We observe that independently of the Lagrangian point, it is verified that $\lambda_2 = - \lambda_1$ and $\lambda_4 = - \lambda_3$. For stable Lagrangian points ($\phi_i = 0, \pi$) all the eigenvalues $\lambda_i$ are purely imaginary, which implies that the solution is given by
\begin{align}
r_1(t) = C_1 \cos(\operatorname{Im}(\lambda_1) t) + C_2 \cos(\operatorname{Im}(\lambda_3) t),\\
\phi_1(t) = C_3 \sin(\operatorname{Im}(\lambda_1) t) + C_4 \sin(\operatorname{Im}(\lambda_3) t),
\end{align}
where the constants $C_i$ are determined in terms of $c_i, V_i$ and the vector $X_p$. For ustable Lagrangian points ($\phi_i = \pi/2, 3\pi/2$) two of the eigenvalues are real and two are imaginary; the solution is
\begin{align}
r_1(t) = \tilde{C}_1 \cos(\operatorname{Im}(\lambda_1) t) + \tilde{C}_2 ( \mathrm{e}^{-\lambda_3 t} + \mathrm{e}^{\lambda_3 t}),\\
\phi_1(t) = \tilde{C}_3 \sin(\operatorname{Im}(\lambda_1) t) +  \tilde{C}_4 ( \mathrm{e}^{-\lambda_3 t} - \mathrm{e}^{\lambda_3 t}),
\end{align}
where it was assumed that $\lambda_1,\lambda_2$ are imaginary and $\lambda_3,\lambda_4$ are real and the constants $\tilde{C}_i$ depend on $c_i, V_i, X_p$.  The two solutions have different limits of validity. Around the stable Lagrange points the solution is valid for all times $t$. On the other hand, around $\phi_i = \pi/2, 3\pi/2$, the unstable points, the solution is valid in a limited range of the time coordinate: the presence of the exponential terms force the values of $r_1$ and $\phi_1$ out of the smallness assumption in which rests the validity of the solution.
\bibliography{references}
\end{document}